\documentclass[preprint,preprintnumbers,amsmath,amssymb,floatfix,nofootinbib]{revtex4}
\usepackage{graphicx}
\newcommand{\sss}{\scriptscriptstyle}
\begin{document}


%
%

\title{Higgs Production in Association with Bottom Quarks at Hadron 
       Colliders}

\author{S.~Dawson}
\email{dawson@quark.phy.bnl.gov}
\affiliation{Physics Department, Brookhaven National Laboratory,
Upton, NY 11973-5000, USA}
\author{C.~B.~Jackson}
\email{jackson@quark.phy.bnl.gov}
\affiliation{Physics Department, Brookhaven National Laboratory,
Upton, NY 11973-5000, USA}
\author{L.~Reina}
\email{reina@hep.fsu.edu}
\affiliation{Physics Department, Florida State University,
Tallahassee, FL 32306-4350, USA}
\author{D.~Wackeroth}
\email{dow@ubpheno.physics.buffalo.edu}
\affiliation{Department of Physics, SUNY at Buffalo,
Buffalo, NY 14260-1500, USA}








\date{\today}

\begin{abstract}

We review the present status of the QCD corrected cross sections and
kinematic distributions for the production of a Higgs boson in
association with bottom quarks at the Fermilab Tevatron and CERN Large Hadron
Collider.  Results are presented for the Minimal Supersymmetric
Standard Model where, for large $\tan\beta$, these production modes
can be greatly enhanced compared to the Standard Model case. The
next-to-leading order QCD results are much less sensitive to the
renormalization and factorization scales than the lowest order
results, but have a significant dependence on the choice of the
renormalization scheme for the bottom quark Yukawa coupling.  We also
investigate the uncertainties coming from the Parton Distribution
Functions and find that these uncertainties can be comparable to the
uncertainties from the remaining scale dependence of the
next-to-leading order results. We present results separately for the
different final states depending on the number of bottom quarks
identified.

\end{abstract}

\maketitle

\section{Introduction}

One of the most exciting prospects of particle physics is the
discovery of the source of Electroweak Symmetry Breaking (EWSB).  In
the Standard Model (SM), a single $SU(2)_L$ complex scalar doublet is
responsible for both gauge boson masses (via the Higgs mechanism) and
fermion masses (via Yukawa interactions).  The only scalar physical
remnant of the process is the Higgs boson ($h$) whose mass is a free
parameter of the model.  In contrast, the Minimal Supersymmetric
Standard Model (MSSM) contains two complex $SU(2)_L$ scalar doublets,
as required to give masses to both up- and down-type quarks and to
cancel gauge anomalies.  Once EWSB occurs in the MSSM, the scalar
sector contains {\it{five}} physical Higgs bosons: a light scalar
$h^0$, a heavy scalar $H^0$, a pseudoscalar $A^0$ and a pair of
charged bosons $H^{\pm}$.  Finding experimental evidence for the
existence of one or more Higgs bosons is, thus, a major goal of
current and future collider experiments.  In addition, measuring the
couplings of the Higgs boson(s) to gauge bosons, leptons and quarks
would be imperative to uncover the true structure of the Higgs sector.

Direct searches for Higgs bosons at LEP2 have placed lower mass limits
on both the SM Higgs ($M_h>$ 114.4~GeV at $95\%$
c.l.)~\cite{Barate:2003sz} and the lightest scalar and pseudoscalar
Higgs bosons of the MSSM ($M_{h^0,A^0} {\stackrel{>}{\scriptstyle
\sim}}$ 93~GeV at $95\%$ c.l.)~\cite{Lephwg2:2004}.  In addition, the
mass of the lightest scalar Higgs boson of the MSSM is also bounded
from above by theory to less than $\sim$ 130~GeV.  Finally, precision
electroweak measurements, which are sensitive to the effects of Higgs
boson loop contributions, imply that the mass of the SM Higgs boson
should be less than $\sim$ 186~GeV at $95\%$
c.l.~\cite{Lepewwg:2004} ($\sim$ 219~GeV when including the LEP-2 direct search limit).  Hence, in both the SM and the MSSM, a
Higgs boson should lie in a mass range which can be explored at the
Large Hadron Collider (LHC)~\cite{Carena:2002es}.

The dominant production mechanism for a SM Higgs boson in hadronic
interactions ($p\bar{p}$ or $pp$) is gluon fusion ($gg\rightarrow h$),
which proceeds predominantly through a loop of top quarks.  Among the
production modes with smaller rates, the associated production with
electroweak gauge bosons ($q\bar{q}\rightarrow Vh$, where $V =
W^{\pm},Z^0$) or with top quark pairs ($q\bar{q},gg\rightarrow
t\bar{t}h$), as well as weak boson fusion ($qq\rightarrow qqh$), play
crucial roles.  The next-to-leading order (NLO) QCD corrections have been calculated for all
of the above processes
~\cite{Han:1991ia,Han:1992hr,Dawson:1991zj,Djouadi:1991tk,Graudenz:1993,Spira:1995,Beenakker:2001rj,Reina:2001sf,Reina:2001bc,Beenakker:2002nc,Dawson:2002tg,Dawson:2003zu},
while for gluon fusion and associated $Vh$ production corrections are
also known at next-to-next-to-leading order (NNLO) in perturbative
QCD~\cite{Harlander:2002wh,Anastasiou:2002yz,Harlander:2003ai,Catani:2001ic,Ravindran:2002dc,Brein:2003wg}.

In the MSSM, the hierarchy of production modes can be significantly
different from that of the SM.  In fact, for large values of
$\tan\beta$ (the ratio of the two MSSM Higgs vacuum expectation
values), the bottom quark Yukawa couplings to both scalar and
pseudoscalar Higgs bosons become strongly enhanced and the production
of Higgs bosons with bottom quarks dominates over all other production
modes.

\begin{figure}[t]
\begin{center}
\includegraphics[scale=0.9]{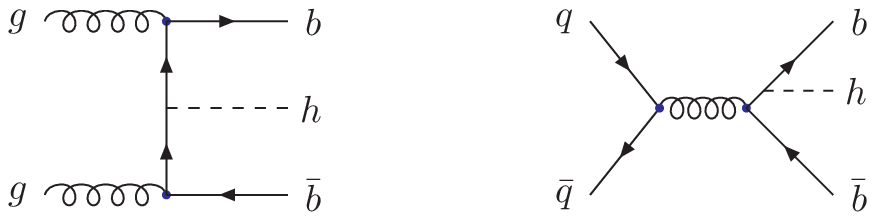}
\vspace*{-0.2cm}
\caption[]{Sample Feynman diagrams for $gg\rightarrow b\bar{b}h$ and 
  $q\bar{q}\rightarrow b\bar{b}h$ production at tree level.}
\label{fg:ggbbh_feyn}
\end{center}
\end{figure}

Recently, the production of Higgs bosons in association with bottom
quarks has received much interest from both the theoretical and
experimental communities~\cite{Campbell:2004,Avto:2005,Acosta:2005bk}.
At tree level, the cross section is almost entirely dominated by the
sub-process $gg\rightarrow b\bar{b}h$,
with only a small contribution from $q\bar{q}\rightarrow b\bar{b}h$,
at both the Tevatron and LHC (see Fig.~\ref{fg:ggbbh_feyn}).  The
theoretical prediction of $b\bar{b}h$ production at hadron colliders,
however, involves several subtle issues, as will be described in
detail in this review, and depends on the number of bottom quarks
identified, or {\it{tagged}}, in the final state.  In case of no or
only one tagged bottom quark there are two approaches available for
calculating the cross sections to $b\bar{b}h$ production, dubbed the
four flavor (4FNS) and five flavor (5FNS) number schemes.  The main
difference between these two approaches is that the 4FNS is a
fixed-order calculation of QCD corrections to the
$gg,q\bar{q}$-induced $b\bar{b}h$ production processes, while in the
5FNS the leading processes are induced by $bg$ ($\bar{b}g$) and
$b\bar{b}$ initial states and large collinear logarithms present at
tree level as well as at each order of QCD corrections are resummed by
using a pertubatively defined bottom quark Parton Distribution
Function (PDF). The two schemes are described in more detail in
Section~\ref{sec:fourfns} and~\ref{sec:fivefns}, respectively.
\begin{figure}[t]
\begin{center}
\includegraphics[scale=0.9]{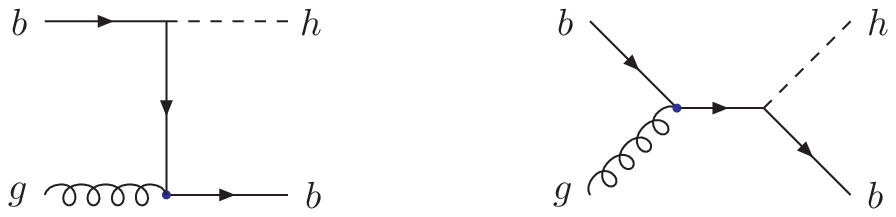}
\vspace*{-0.2cm}
\caption[]{Tree level Feynman diagram for $bg\rightarrow bh$ in the 5FNS.}
\label{fg:bghb_feyn}
\end{center}
\end{figure}
\begin{figure}[t]
\begin{center}
\includegraphics[scale=0.9]{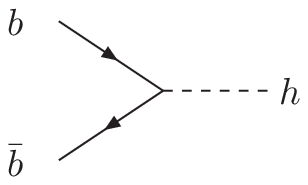}
\vspace*{-0.2cm}
\caption[]{Tree level Feynman diagram for $b\bar{b}\rightarrow h$ in the 5FNS.}
\label{fg:bbh_feyn}
\end{center}
\end{figure}

For purposes of differentiating the final states, we introduce the
following terminology.  {\it{Exclusive}} $b\bar{b}h$ production refers
to the case when both bottom quark jets are tagged (or are at high
transverse momentum, $p_T$).  Note that, in this case, the computation
of the cross section solely relies on the 4FNS.  {\it{Semi-inclusive}}
production (denoted by $b(\bar{b})h$) pertains to the case when
{\it{at least}} one bottom quark is at high $p_T$.  Here, the dominant
leading-order (LO) process for the 4FNS (5FNS) is $gg\rightarrow
b(\bar{b})h$ ($bg\rightarrow bh$ or $\bar{b}g\rightarrow \bar{b}h$,
see Fig.~\ref{fg:bghb_feyn}).  Finally, the {\it{inclusive}}
production process (denoted by $(b\bar{b})h$) involves the case where
no bottom quark jets are tagged.  The dominant LO process for this
mode is $gg
\rightarrow (b\bar{b})h$ ($b\bar{b}\rightarrow h$, see Fig.~\ref{fg:bbh_feyn})
in the 4FNS (5FNS).  It should be mentioned that, although the cross
section for the inclusive production mode dominates over the less
inclusive modes by one to two orders of magnitude, tagging bottom
quark jets greatly reduces backgrounds, making the less inclusive
modes experimentally more appealing.  Requiring one or two high-$p_T$
bottom quark jets in the final state also ensures that the detected
Higgs boson has been radiated off a bottom or anti-bottom quark and
the corresponding cross section is unambiguously proportional to the
bottom quark Yukawa coupling.

QCD corrected cross sections are available for all three final states
and the inclusive and semi-inclusive production processes have been
computed in both the 4FNS and 5FNS.  For the inclusive case, the NLO
QCD corrected 4FNS~\cite{Dittmaier:2003ej,Campbell:2004} and the NNLO
QCD corrected 5FNS cross sections~\cite{Harlander:2003ai} have been
compared and are found to be in good agreement within theoretical
uncertainties.  The NLO predictions of the semi-inclusive cross
sections for the 4FNS~\cite{Dittmaier:2003ej,Dawson:2004tl} and
5FNS~\cite{Campbell:2002zm} have also been extensively compared~\cite{Campbell:2004,Dawson:2004tl} and
the agreement between the two is spectacular.  The compatibility of
these two seemingly different calculational schemes in the prediction
of inclusive and semi-inclusive Higgs boson production rates is indeed
a powerful check of the theory. The 4FNS and 5FNS represent different
perturbative expansions of the same physical observables, and
therefore should agree at sufficiently high order. Indeed, we see that
considering the first (second) order of corrections already brings
agreement between the two schemes within the respective theoretical
uncertainties.  Finally, two independent calculations of the NLO QCD
corrections for the exclusive mode have been compared and agreement
has been found~\cite{Dittmaier:2003ej,Dawson:2003ex,Campbell:2004}.

The above discussion for the production of a {\it{scalar}} Higgs boson
with bottom quarks applies equally well to the production of a
{\it{pseudoscalar}} Higgs boson.  In fact, for massless bottom quarks,
the predictions for $b\bar{b}A^0$ would be identical to those of
$b\bar{b}h^0/H^0$ given a rescaling of the Yukawa couplings.  For massive
bottom quarks, however, the situation becomes complicated due to the
$\gamma_5$ appearing in the $b\bar{b}A^0$ Yukawa coupling.  The
$\gamma_5$ Dirac matrix is intrinsically a four-dimensional object and
care must be taken in its treatment when dimensionally regularizing
the calculation.  In any case, bottom quark mass effects are expected
to be small, i.e. $\mathcal{O}$$(\frac{m_b^2}{M_{h}^2})$, and predictions for
$b\bar{b}h^0/H^0$ are good indicators for $b\bar{b}A^0$ production
(modulus Yukawa couplings).

The remainder of this paper is organized as follows.  In
Section~\ref{sec:theory}, we discuss the framework of the calculations
in the four- and five-flavor-number schemes in detail.  Given the
importance it plays in the prediction of the cross section through the
Yukawa coupling, we also discuss different renormalization schemes of
the bottom quark mass.  Section~\ref{sec:MSSMcoups} details the
SUSY-corrected Yukawa couplings obtained from the program {\sc
FeynHiggs}~\cite{Heinemeyer:2001qd,Frank:2002qa,Heinemeyer:2004ms}
that were used in this study to produce results for MSSM Higgs bosons.
In Section~\ref{sec:numbers}, we review the results for inclusive,
semi-inclusive and exclusive Higgs boson production with bottom
quarks.  In Section~\ref{sec:pdfs}, we investigate the uncertainties
arising from the PDFs for the semi-inclusive $b\bar{b}h$
production process.  These uncertainties are calculated using the
algorithm developed by the CTEQ collaboration~\cite{CTEQ:2002}.
Finally, we summarize and conclude in Section~\ref{sec:summary}.

\section{Theoretical Framework}
\label{sec:theory}
\subsection{Five Flavor Number Scheme}
\label{sec:fivefns}

Final state bottom quarks are identified imposing cuts on their
transverse momentum ($p_T\!>\!p_T^{min}$) and pseudorapidity
($|\eta|\!<\!\eta^{max}$). On the other hand, when a final state
bottom quark is not identified, as in the inclusive and semi-inclusive
$b\bar{b}h$ production processes, the corresponding integration over
its phase space may give rise to large logarithms of the form:
\begin{equation}
\label{eq:lambda_b}
\Lambda_b \equiv \ln\left(\frac{\mu^2}{m_b^2}\right)\,\,\,,
\end{equation}
where $m_b$ and $\mu$ represent the lower and upper bounds on the
integration over the transverse momentum, $p_T$, of the final state
bottom quark. This happens when the final state bottom quark is
directly originating from the $g\rightarrow b\bar{b}$ splitting of an
initial state gluon, and corresponds to the collinear singularity that
would be present in the case of a gluon splitting into massless
quarks.  The scale $\mu$ is typically of $\mathcal{O}(M_h)$ and
therefore, due to the smallness of the bottom quark mass, these
logarithms can be quite large.  Additionally, the same logarithms
appear at every order in the perturbative expansion of the cross
section in the strong coupling, $\alpha_s$, due to recursive gluon
emission from internal bottom quark lines.  These logarithms could
severely hinder the convergence of the perturbative expansion of the
total and differential cross sections.  In the 5FNS the convergence is
improved by introducing a perturbatively defined bottom quark
PDF~\cite{Barnett:1988,Olness:1988,Dicus:1989}, defined at lowest
order in $\alpha_s$ as
\begin{equation}
\label{eq:b_pdf_lo}
b(x,\mu)=\frac{\alpha_s(\mu)}{2\pi}\Lambda_b \int_x^1
\frac{dy}{y} P_{qg}\left(\frac{x}{y}\right)g(y,\mu)\,\,\,,
\end{equation}
\noindent where $g(y,\mu)$ is the (experimentally-measured) gluon PDF and 
$P_{qg}$ is the Altarelli-Parisi splitting function for $g\rightarrow
q\bar{q}$ given by
\begin{equation}
\label{eq:ap_qg}
P_{qg}(z) = \frac{1}{2}[z^2 + (1-z)^2]\,\,\,.
\end{equation}
Subsequently, the $\Lambda_b$ logarithms are resummed through the
DGLAP evolution
equation~\cite{Gribov:1972,Altarelli:1977zs,Dokshitzer:1977}, such
that contributions proportional to $(\alpha_s\Lambda_b)^n$ can be
absorbed, to all orders in $n$, into a leading logarithms bottom quark
PDF, while subleading logarithms are recursively resummed when higher
order corrections are considered in the DGLAP equation. Since
$\alpha_s\Lambda_b$ is not a small expansion parameter, the use of a
bottom quark PDF should improve the stability of the total and
differential cross sections for inclusive and semi-inclusive Higgs
boson production with bottom quarks.

The 5FNS is based on the approximation that any \emph{spectator}
bottom quarks (i.e., bottom quarks which are not tagged in the final
state) are at small transverse momentum, since this is the region
where the $\Lambda_b$ logarithms dominate over other $p_T$-dependent
contributions.  At lowest order, the spectator quarks are produced
with zero $p_T$, and a transverse momentum spectrum for the outgoing
bottom quarks is generated at higher
orders~\cite{Barnett:1988,Olness:1988}.

With the use of a bottom quark PDF, the 5FNS effectively reorders the
perturbative expansion to be one in $\alpha_s$ {\it{and}}
$\Lambda_b^{-1}$.  To see how this works, let us consider the
perturbative expansion of the inclusive process $b\bar{b}\rightarrow
h$ (Fig.~\ref{fg:bbh_feyn}), which is at tree level of order
$\alpha_s^2\Lambda_b^2$.  At NLO, the virtual and real corrections to
the tree level process make contributions of
$\mathcal{O}(\alpha_s^3\Lambda_b^2)$.  However, at NLO, we must also
consider the contribution from $bg\rightarrow bh$ where the final
state bottom quark is at high $p_T$.  This process makes a
contribution of order $\alpha_s^2\Lambda_b$ and is, thus, a correction
of $\mathcal{O}(\Lambda_b^{-1})$ to the tree level cross section.
Similarly, at NNLO, among the myriad of radiative corrections of
$\mathcal{O}(\alpha_s^4\Lambda_b^2)$, we must also include the
contribution from the process $gg\rightarrow b\bar{b}h$, where both
$b$ and $\bar{b}$ are at high $p_T$.  The contribution from these
diagrams are of order $\alpha_s^2$, and are, thus,
$\mathcal{O}(\Lambda_b^{-2})$ (or NNLO) corrections to the tree level
process $b\bar{b}\rightarrow h$~\cite{Dicus:1989,Dicus:1998hs}.

The above discussion for $b\bar{b}\rightarrow h$ also applies to the
perturbative expansion of $bg\rightarrow bh$.  In this case, the tree
level process is of order $\alpha_s^2 \Lambda_b$ and the contribution
from $gg\rightarrow b\bar{b}h$ is a NLO correction of
${\cal{O}}(\Lambda_b^{-1})$~\cite{Campbell:2002zm}.

Finally, we observe that, in the existing 5FNS calculations, all bottom
quark masses other than the one appearing in the Yukawa coupling are
set to zero.

\subsection{Four Flavor Number Scheme}
\label{sec:fourfns}

In the 4FNS, the initial state quarks are constrained to be the four
lightest quarks, i.e. there are no bottom quarks in the initial state.
No kinematic approximations are made and the cross section for
$p\bar{p}(pp)\rightarrow b\bar{b}h$ is computed at fixed order in QCD
without resumming higher-order collinear logarithms. Moreover, the
bottom quark mass is always considered to be non-zero.  Similarly to
the 5FNS, for the less inclusive final states discussed above
($b(\bar{b})h$ or $b\bar{b}h$), identification cuts are imposed on the
final state bottom quark(s) which constrain its transverse momentum
and pseudorapidity. Applying these cuts to a final state bottom quark
eliminates large logarithms that may appear from gluon splitting and
it ensures that the bottom quark jets can be tagged
experimentally. While the inclusive and semi-inclusive production of a
Higgs boson with bottom quarks can be calculated in both the 4FNS and
5FNS, the exclusive production can only be calculated in the 4FNS,
since both final state bottom quarks are tagged.

The NLO QCD corrections to the hadronic process $p\bar{p}(pp)
\rightarrow b\bar{b}h$ in the 4FNS consist of calculating the
$\mathcal{O}(\alpha_s)$ virtual and real QCD corrections to the tree
level (partonic) processes $q\bar{q},gg \rightarrow
b\bar{b}h$~\cite{Dittmaier:2003ej,Dawson:2003ex}. We note that the NLO
calculation of $p\bar{p}(pp)\rightarrow b\bar{b}h$ is identical to the
calculation of the NLO QCD corrections to $p\bar{p}(pp)\rightarrow
t\bar{t}h$~\cite{Beenakker:2001rj,Reina:2001sf,Reina:2001bc,Beenakker:2002nc,Dawson:2002tg},
with the global interchange of the top quark and bottom quark masses
($m_t\leftrightarrow m_b$).

\subsection{Definition of the bottom quark mass}
\label{sec:renorm}

One potential source of theoretical uncertainty in the calculation of
$p\bar{p}(pp)\rightarrow b\bar{b}h$ involves the renormalization of the
bottom quark mass.  The bottom quark mass counterterm has to be used
twice in the renormalization of the calculation: once to renormalize
the bottom quark mass appearing in internal propagators and once to
renormalize the SM bottom quark Yukawa coupling $g_{b\bar{b}h}^{\sss
S\!M}\!=\!m_b/v$, where $v=(\sqrt{2} G_F)^{-1/2}=246$~GeV.  Indeed, if
one only considers QCD corrections, the counterterm for the bottom
quark Yukawa coupling coincides with the counterterm for the bottom
quark mass, since the vacuum expectation value $v$ is not renormalized
at 1-loop in QCD.

The renormalization scheme dependence has been studied in
Ref.~\cite{Dawson:2003ex}, in the 4FNS, for the case of the exclusive
$b\bar bh$ production. Two schemes for the renormalization of the
bottom quark mass have been chosen: the $\overline{MS}$ scheme and an
on-shell ($OS$) scheme.  The two are perturbatively consistent at NLO
with the only difference being at higher orders and, thus, part of the
theoretical uncertainty of the NLO calculation.  The main reason for
investigating this renormalization scheme dependence is the large
sensitivity of the $\overline{MS}$ bottom quark mass and the prominent
role it plays in the $b\bar{b}h$ production cross section through the
overall bottom quark Yukawa coupling.  A summary of the results of
this investigation are presented in Section~\ref{subsec:2btag}.

The benefit of using the $\overline{MS}$ scheme for the Yukawa
coupling is that it potentially gives control over higher-order
corrections beyond the 1-loop corrections. This is often reflected in
the better perturbative behavior of physical observables calculated
using the $\overline{MS}$ bottom quark Yukawa coupling.  Therefore,
unless otherwise stated all the results presented in this review are
given using the $\overline{MS}$ scheme for the bottom quark mass and
Yukawa coupling.  This is accomplished by replacing the
$\overline{MS}$ bottom quark mass in the Yukawa coupling with the
corresponding 1-loop and 2-loop renormalization group improved
$\overline{MS}$ masses:
\begin{equation}
\label{eq:mb_mu_lo}
{\overline m}_b(\mu)_{1l}=m_b\left[\frac{\alpha_s(\mu)}{\alpha_s(m_b)}
\right]^{c_0/b_0}\,\,\,,
\end{equation}
\begin{equation}
\label{eq:mb_mu_nlo}
{\overline m}_b(\mu)_{2l}=m_b\left[\frac{\alpha_s(\mu)}{\alpha_s(m_b)}
\right]^{c_0/b_0}
\left[ 1+\frac{c_0}{b_0}(c_1-b_1)\frac{\alpha_s(\mu)-\alpha_s(m_b)}{\pi}
\right] \left(1-\frac{4}{3}\frac{\alpha_s(m_b)}{\pi}\right)\,\,\,,
\end{equation}
\noindent where we take the bottom quark pole mass to be $m_b\!=\!4.62$~GeV
and
\begin{eqnarray}
\label{eq:b0_b1_c0_c1}
b_0&=\frac{1}{4\pi}(\frac{11}{3}N-\frac{2}{3} n_{lf})\,\,\,,\,\,\,
& c_0=\frac{1}{\pi}\,\,\,,\nonumber \\
b_1&=\frac{1}{2\pi}\frac{51N-19 n_{lf}}{11N-2 n_{lf}}\,\,\,,
& c_1=\frac{1}{72\pi}(101N-10 n_{lf})
\end{eqnarray}
\noindent are the one and two loop coefficients of the QCD $\beta$ function
and mass anomalous dimension $\gamma_m$, while $N=3$ is the number of colors
and $n_{lf}=5$ is the number of light flavors.

\section{Radiatively-corrected MSSM bottom Yukawa Coupling}
\label{sec:MSSMcoups}

\begin{figure}[t]
\begin{center}
\includegraphics[scale=0.7]{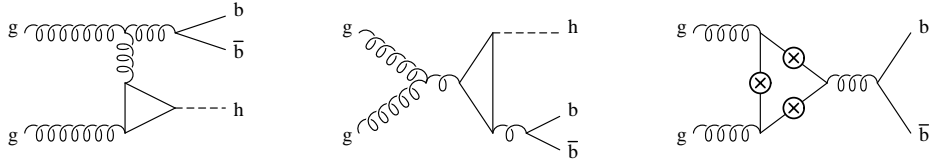}
\vspace*{-0.2cm}
\caption[]{Sample Feynman diagrams for the closed top quark loop
contribution to $gg\rightarrow b\bar bh$.  The circled cross denotes
all possible insertion of the final state Higgs boson leg, each
insertion corresponding to a different diagram. }
\label{fg:bbhtriangle}
\end{center}
\end{figure}

So far, most of the results for $b\bar{b}h$ production have been
obtained in the SM, in order to simplify the comparison of different
calculations and approaches (for a review see, e.g.,
Ref.~\cite{Campbell:2004}).  As stated earlier, though, these
processes are relevant discovery modes at hadron colliders only for
SUSY models with large $\tan\beta$. Therefore, we will present most of
the following results for the case of the MSSM with large
$\tan\beta$. The prediction for the MSSM case can easily be derived
from the SM results by rescaling the Yukawa couplings.  In the 4FNS,
due to the existence of diagrams where the Higgs is radiated off a
closed top quark loop, some care must be taken when performing this
rescaling. In this review, all 4FNS results for the neutral MSSM Higgs
bosons, are obtained from the SM results by carefully rescaling the
contributions {\it{with}} ($\sigma_{\sss S\!M}$) and {\it{without}}
top quark loops ($\sigma_{\sss S\!M} - \sigma_{\sss S\!M}^t$) as:
\begin{eqnarray}
\label{eq:sigma_mssm_rescaled}
\sigma_{\sss M\!S\!S\!M}&=&
\left(\frac{g_{b\bar{b}h}^{\sss M\!S\!S\!M}}{g_{b\bar{b}h}^{\sss S\!M}}\right)^2
\left(\sigma_{\sss S\!M}-\sigma_{\sss S\!M}^t\right)
+\left(\frac{g_{t\bar{t}h}^{\sss M\!S\!S\!M}g_{b\bar{b}h}^{\sss M\!S\!S\!M}}
{g_{t\bar{t}h}^{\sss S\!M} g_{b\bar bh}^{\sss S\!M}}\right)
\sigma_{\sss S\!M}^t\,\,\,.
\label{eq:rescale}
\end{eqnarray}
Examples of the closed top quark loop contributions described by
$\sigma_{\sss S\!M}^t$ are shown in Fig.~\ref{fg:bbhtriangle}. Note
that in the 5FNS, as implemented in the program {\sc
MCFM}~\cite{MCFM:2004}, the corresponding contributions to
$\sigma_{\sss S\!M}^t$ are considered to be zero. While this could
affect the comparison in the SM, in SUSY models with large bottom
quark Yukawa couplings (and consequently small top quark Yukawa
couplings) the contributions of the closed top quark diagrams is very
small, and $\sigma_{\sss S\!M}^t$ does not play a significant role.

In the MSSM the bottom and top quark Yukawa couplings to the scalar,
neutral Higgs bosons are given by:
\renewcommand{\arraystretch}{2}
\begin{equation}
\begin{array}{lclrcr}
g_{b\bar{b}h^0}^{\sss M\!S\!S\!M} & = & 
-\displaystyle{\frac{\sin\alpha}{\cos \beta}}g_{b\bar{b}h}^{\sss S\!M}\,\,\,,\,\,\,& 
g_{t\bar{t}h^0}^{\sss M\!S\!S\!M} & = &
\displaystyle{\frac{\cos\alpha}{\sin\beta}}g_{t\bar{t}h}^{\sss S\!M}\,\,\,, \\
g_{b\bar{b}H^0}^{\sss M\!S\!S\!M} & = & 
\displaystyle{\frac{\cos\alpha}{\cos\beta}}g_{b\bar{b}h}^{\sss S\!M}\,\,\,,\,\,\,&
g_{t\bar{t}H^0}^{\sss M\!S\!S\!M} & = & 
\displaystyle{\frac{\sin\alpha}{\sin\beta}}g_{t\bar{t}h}^{\sss S\!M}\,\,\,,
\end{array}
\end{equation}
\renewcommand{\arraystretch}{1} where $g_{b\bar{b}h}^{\sss S\!M}$ and
$g_{t\bar{t}h}^{\sss S\!M}$ are the SM bottom and top quark Yukawa
couplings, $h^0$ and $H^0$ are the lighter and heavier neutral scalars
of the MSSM, and $\alpha$ is the angle which diagonalizes the neutral
scalar Higgs mass matrix~\cite{Gunion:1989we}.
The dominant SUSY radiative corrections to $b\bar bh$ production can
be taken into account by including the SUSY corrections to the $b\bar
bh$ vertex only, i.e. by replacing the tree level Yukawa couplings by
the radiative corrected ones. We follow the treatment of the program
{\sc FeynHiggs}~\cite{FeynHiggs:2005} and take into account the
leading, $\tan\beta$ enhanced, radiative corrections that are
generated by gluino-sbottom and chargino-stop loops, as
follows~\cite{Carena:1999py}:
\begin{eqnarray}
\label{eq:bbh_rad}
g_{b\bar{b}h^0}^{\sss M\!S\!S\!M} &=& 
-g_{b\bar{b}h}^{\sss S\!M}\frac{1}{1+\Delta_b}
\left[\frac{\sin\alpha}{\cos\beta}-\Delta_b\frac{\cos\alpha}{\sin\beta}\right]\,\,\,,
\nonumber\\
g_{b\bar{b}H^0}^{\sss M\!S\!S\!M} &=& \,\,\,\,g_{b\bar{b}h}^{\sss S\!M}\frac{1}{1+\Delta_b} 
\left[\frac{\cos\alpha}{\cos\beta}+\Delta_b \frac{\sin\alpha}{\sin\beta}\right]\,\,\,,
\end{eqnarray}
\noindent with
\begin{eqnarray}
\Delta_b &=& \mu\tan\beta \left[\frac{2 \alpha_s(m_t)}{3\pi } M_{\tilde g} 
I(m_{\tilde b_1},m_{\tilde b_2},m_{\tilde g})+\left(\frac{h_t}{4\pi}\right)^2 
A_t I(m_{\tilde t_1},m_{\tilde t_2},\mu)\right] \,\,\,,
\end{eqnarray}
where $h_t=\overline{m}_t(m_t)/\sqrt{v_1^2+v_2^2}\approx
\overline{m}_t(m_t)/174$~GeV, $m_{\tilde b_{1,2}}, m_{\tilde t_{1,2}},
M_{\tilde g}$ denote the sbottom, stop, and gluino masses,
respectively, $A_t$ is the stop and $\mu$ the Higgs mixing parameter.
The strong coupling constant and the running top quark mass,
$\overline{m}_t$, are evaluated at the on-shell top mass,
$m_t\!=\!174$~GeV. The vertex function $I$ is given by
\begin{eqnarray}
\label{eq:i_function}
I(a,b,c)&=&\frac{[a^2 b^2 \ln(\frac{a^2}{b^2})+b^2 c^2 \ln(\frac{b^2}{c^2})+
a^2 c^2 \ln(\frac{c^2}{a^2})]}{(a^2-b^2)(b^2-c^2)(a^2-c^2)}\,\,\,.
\end{eqnarray}
In Tables~\ref{tab:ratiolight} and \ref{tab:ratioheavy} we provide the
ratios of the SM and MSSM bottom and top quark Yukawa couplings at
$\tan\beta\!=\!40$, as provided by {\sc FeynHiggs}, where we assumed
CP conserving couplings~\footnote{MSSM Yukawa couplings are also
provided by the programs HDECAY~\cite{Djouadi:1997yw} and {\sc
CPSuperH}~\cite{Lee:2003nt,CPSuperH}.}. In the following, we will use
these ratios to derive the MSSM cross sections from the SM results as
described in Eq.~(\ref{eq:rescale}).  Using {\sc FeynHiggs}, the MSSM
Higgs boson masses and the mixing angle $\alpha$ have been computed up
to two-loop order.  As expected, for this choice of MSSM input
parameters the bottom Yukawa coupling is strongly enhanced and the top
Yukawa coupling is suppressed, resulting in a MSSM $b\bar{b}h$ cross
section that is about three orders of magnitude larger than the SM
cross section.

\begin{table}[t]
{\begin{tabular}{@{}ccccc@{}}\toprule
\multicolumn{5}{c}{$\tan\beta=40$} \\ \hline $M_{h^0}$ [GeV] & 100 &
110 & 120 & 130 \\ \hline $\alpha$ [rad] & -1.5063 & -1.4716 & -1.3798
& -0.7150 \\ $g_{b\bar{b}h^0}^{\sss M\!S\!S\!M}/g_{b\bar{b}h}^{\sss
S\!M}$&33.913 & 33.823 & 33.387 &22.390 \\ $g_{t\bar{t}h^0}^{\sss
M\!S\!S\!M}/g_{t\bar{t}h}^{\sss S\!M}$& 0.0645 & 0.0991 & 0.1899 &0.7553 \\
\botrule
\end{tabular}}
\caption{Ratios of the MSSM and SM bottom and top quark Yukawa couplings
to a light neutral scalar, along with the corresponding values for
$\alpha$ as computed by {\sc
FeynHiggs}~\protect\cite{FeynHiggs:2005}. The genuine SUSY input
parameters are chosen as follows: all soft-SUSY breaking masses for
squark doublets and singlets are set to $M_{SUSY}\!=\!1$~TeV,
$M_{\tilde g}\!=\!1$~TeV, $A_b\!=\!A_t\!=\!2$~TeV, and
$\mu\!=\!M_2\!=\!200$~GeV. For this choice we find $\Delta_b\!=\!0.178$.}
\label{tab:ratiolight}
\end{table}

\begin{table}[t]
{\begin{tabular}{@{}cccccccc@{}}
\toprule \multicolumn{8}{c}{$\tan\beta=40$} \\ \hline $M_{H^0}$ [GeV]
& 120 & 200 & 300 & 400 & 500 & 600 & 800 \\ \hline $\alpha$ [rad]
&-0.3315 & -0.0454 & -0.0318 & -0.0282 & -0.0266 & -0.0258 & -0.0251
\\ $g_{b\bar{b}H^0}^{\sss M\!S\!S\!M}/g_{b\bar{b}h}^{\sss S\!M}$&25.787 &
27.338 &27.356 & 27.360 &27.362 & 27.363 & 27.364 \\
$g_{t\bar{t}H^0}^{\sss M\!S\!S\!M}/g_{t\bar{t}h}^{\sss S\!M}$&-0.3255 &-0.0454
& -0.0318 &-0.0282 &-0.0266 & -0.0258 & -0.0251\\ \botrule
\end{tabular}}
\caption{Ratios of the MSSM and SM bottom and top quark Yukawa couplings
to a heavy neutral scalar, along with the corresponding values for
$\alpha$ as computed by {\sc FeynHiggs}~\protect\cite{FeynHiggs:2005}.
The genuine SUSY input parameters are chosen as follows: all soft-SUSY
breaking masses for squark doublets and singlets are set to
$M_{SUSY}\!=\!1$~TeV, $M_{\tilde g}\!=\!1$~TeV,
$A_b\!=\!A_t\!=\!25$~GeV, and $\mu\!=\!M_2\!=\!1000$~GeV. For this
choice we find $\Delta_b\!=\!0.461$.}  
\label{tab:ratioheavy}
\end{table}

\section{Numerical Results}
\label{sec:numbers}
If not stated otherwise, the numerical results are obtained using
CTEQ6M parton distribution functions for the calculation of the NLO
cross section, and CTEQ6L parton distribution functions for the
calculation of the lowest order cross sections~\cite{Lai:1999wy}. The
NLO (LO) cross section is evaluated using the 2-loop (1-loop)
evolution of $\alpha_s(\mu)$ with $\alpha_s^{NLO}(M_Z)=0.118$.  For
the exclusive and semi-inclusive channels ($b{\overline b}h$ and
$bh+\bar{b}h$ production), it is required that the final state bottom
quarks have $p_T\!>\!20~$GeV and pseudorapidity
$\mid\!\eta\!\mid<\!2.0$ for the Tevatron and $\mid\!\eta\!\mid<\!2.5$
for the LHC. In the NLO real gluon emission contributions, the final
state gluon and bottom quarks are considered as separate particles
only if their separation in the pseudorapidity-azimuthal angle plane,
$\Delta R\!=\!\sqrt{(\Delta\eta)^2+(\Delta\phi)^2}$, is larger than
$0.4$. For smaller values of $\Delta R$, the four momentum vectors of
the two particles are combined into an effective bottom/anti-bottom
quark momentum four-vector.

In the following numerical discussion, the SM NLO QCD corrected cross
sections for semi-inclusive and exclusive $b\bar bh$ production in the
4FNS are taken from Refs.~\cite{Dawson:2004tl}
and~\cite{Dawson:2003ex}, respectively.  The 5FNS SM results for
semi-inclusive $b(\bar b) h$ production have been produced with the
program {\sc MCFM}~\cite{MCFM:2004} and for the inclusive case have
been taken from Ref.~\cite{Harlander:2003ai}.

\subsection{Results for inclusive $(b\bar b)h$ production}
\label{subsec:0btag}
\begin{figure}[t]
\begin{center}
\begin{tabular}{rl}
\includegraphics[bb=150 500 430 700,scale=0.70]{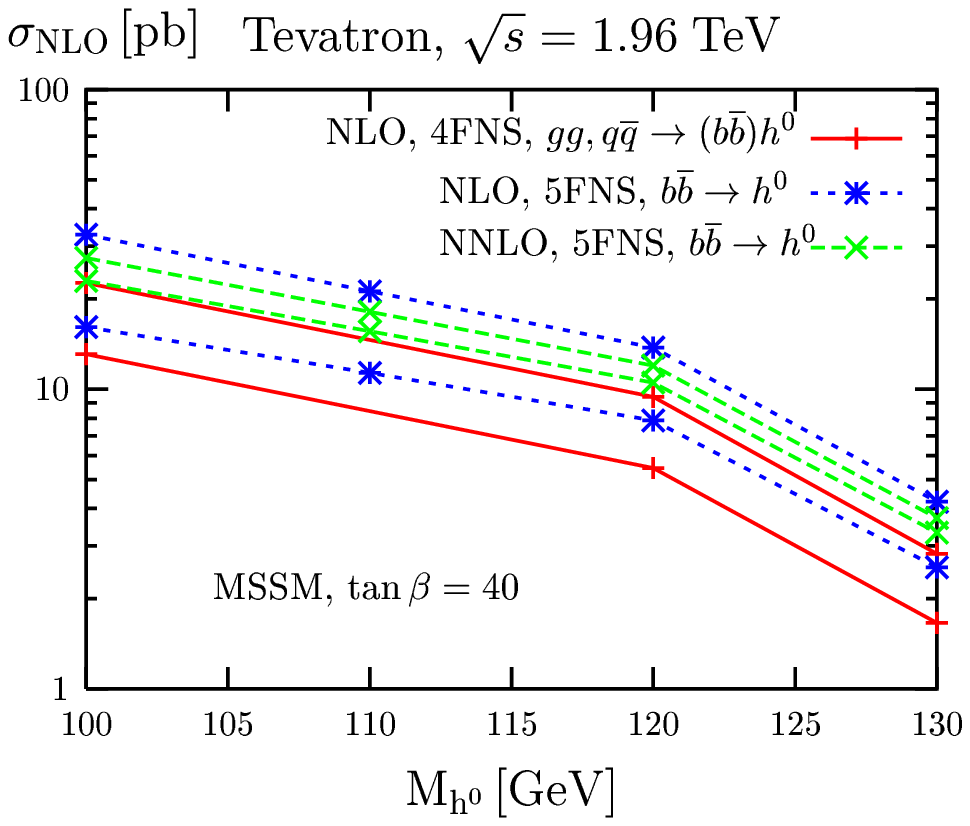} & 
\includegraphics[bb=150 500 430 700,scale=0.70]{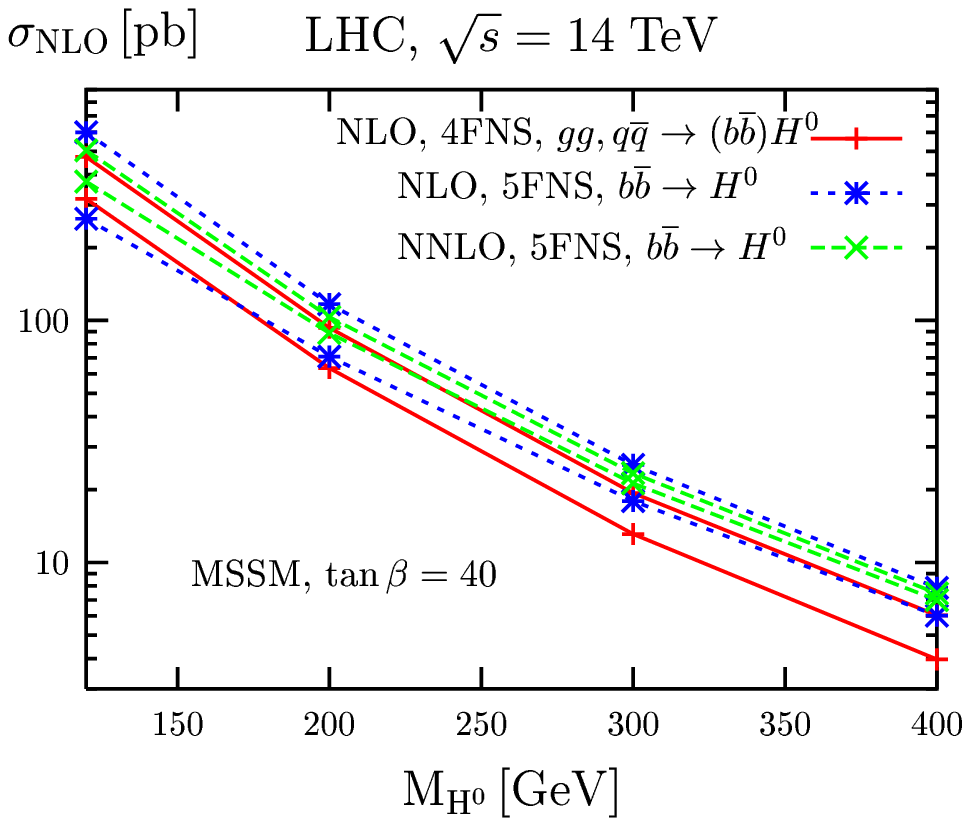}
\end{tabular}
\vspace*{8pt}
\caption[]{Total cross sections for $pp,p\bar p\rightarrow (b\bar{b})h$
  ($h=h^0,H^0$) in the MSSM with no bottom quark jet identified in the
  final state in the 4FNS (at NLO) and 5FNS (at NLO and NNLO) as a
  function of the light and heavy MSSM Higgs boson masses, at both the
  Tevatron and the LHC.  The error bands have been obtained by varying
  the renormalization and factorization scales as
  described in the text.}
\label{fg:mhdep0b}
\end{center}
\end{figure}

If the outgoing bottom quarks cannot be observed then the dominant
MSSM Higgs production process at large $\tan\beta$ is $gg\rightarrow
(b\bar{b})h$ (4FNS) or $b\bar{b}\rightarrow h$ (5FNS). At the LHC,
this process can be identified by the decays to $\mu^+\mu^-$ and
$\tau^+\tau^-$ for the heavy Higgs bosons, $H^0$ and $A^0$, of the MSSM
where the Higgs couplings are enhanced at large $\tan\beta$. Recently,
also at the Tevatron this process, with $h\rightarrow \tau^+\tau^-$,
has been used to search for the neutral MSSM Higgs
boson~\cite{Acosta:2005bk}.  In the 5FNS this process has been
computed to NNLO~\cite{Harlander:2003ai}, and has only a small
renormalization and factorization scale dependence.  In the 4FNS, the
production processes $gg,q\bar q\rightarrow (b {\overline b}) h$,
where the outgoing bottom quarks are not observed, are known at
NLO~\cite{Dittmaier:2003ej,Campbell:2004} and have been recalculated
for the purpose of this review.  A comparison of the total production
rates in the 4FNS and 5FNS within the MSSM for $\tan\beta=40$ is shown
in Fig.~\ref{fg:mhdep0b}. The bands represent the theoretical
uncertainty due to the residual scale dependence at NLO (NNLO). In the
4FNS they have been obtained by varying the renormalization ($\mu_r$)
and factorization ($\mu_f$) scales independently from $\mu_0/4$ to
$\mu_0$, where $\mu_0\!=\!m_b+M_h/2$. In the 5FNS, the renormalization
scale is fixed to $\mu_r=M_h$ and the factorization scale is varied
around the central scale $\mu_f=M_h/4$ from $0.1 M_h$ to $0.7
M_h$. There is good agreement between the results of the two schemes
within their respective scale uncertainties, although the 5FNS at NNLO
is slightly higher than the 4FNS for all values of the Higgs boson
masses.

\subsection{Results for semi-inclusive $b(\bar{b})h$ production}
\label{subsec:1btag}
\begin{figure}[t]
\begin{center}
\begin{tabular}{rl}
\includegraphics[bb=40 30 520 430,scale=0.4]{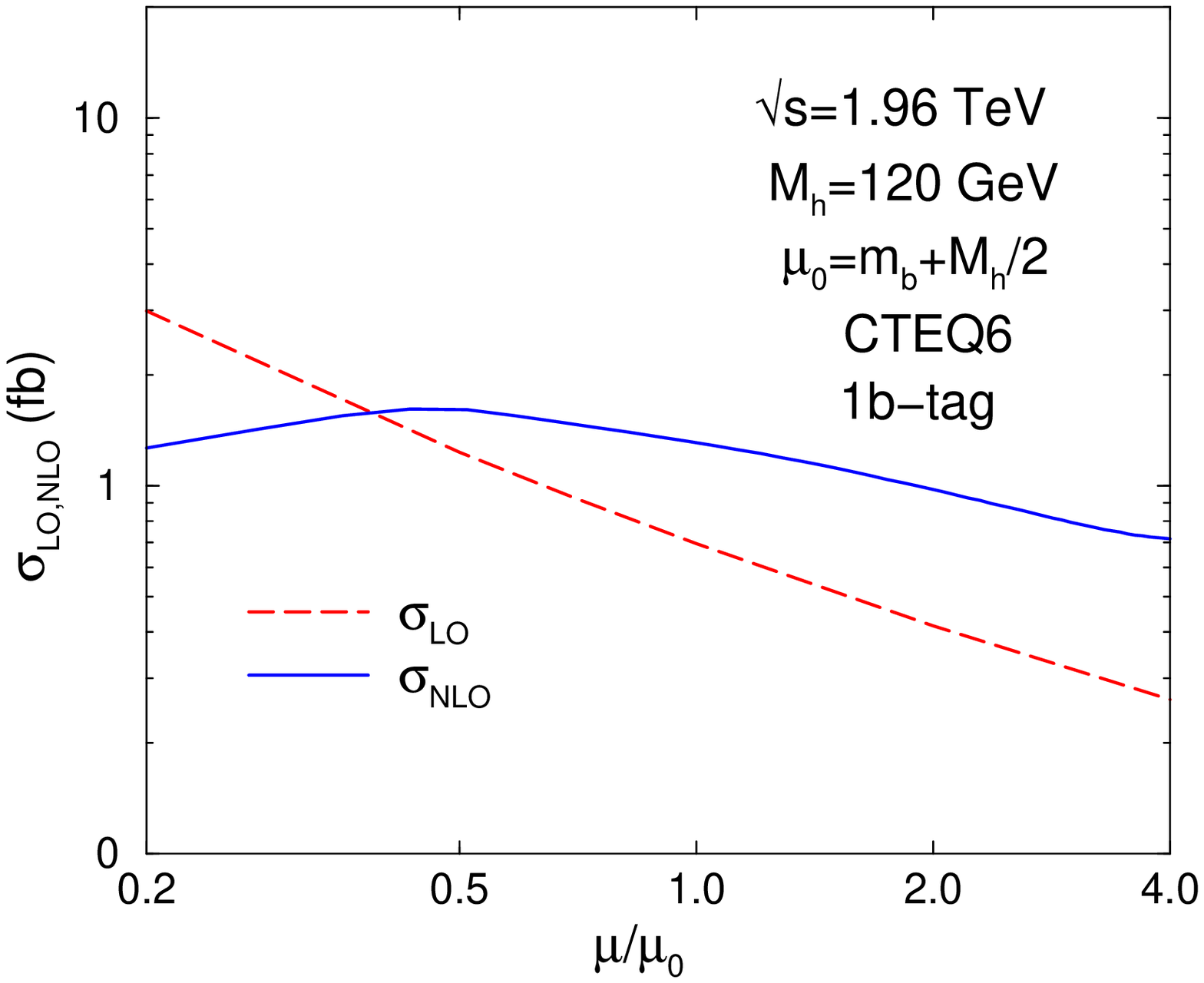} &
\includegraphics[bb=40 30 520 430,scale=0.4]{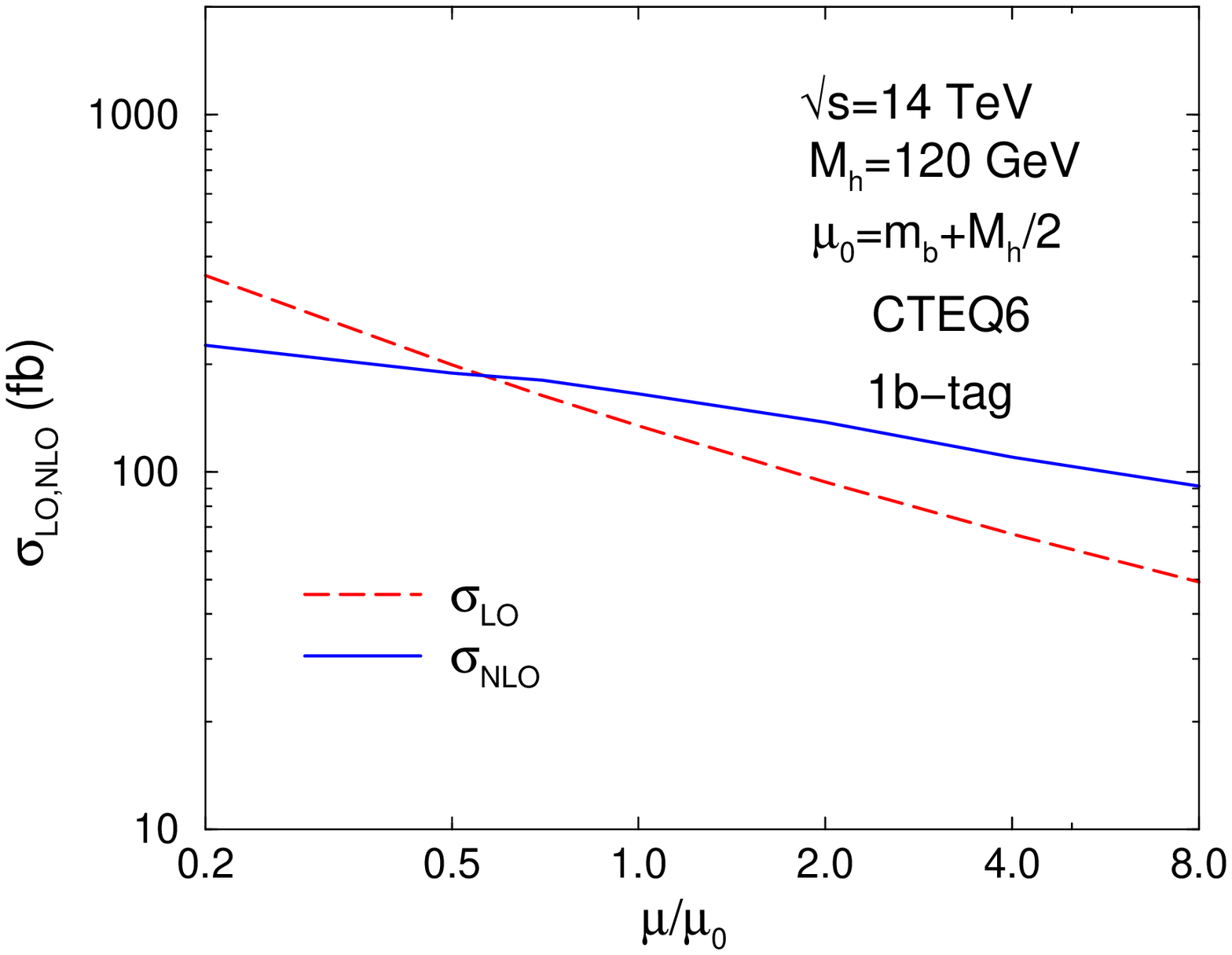}
\end{tabular}
\vspace*{-0.4cm}
\caption[]{Total LO and NLO cross sections in the SM for 
$pp,p\bar{p}\rightarrow b(\bar{b}) h$ production in the 4FNS as a
function of $\mu\!=\!\mu_r\!=\!\mu_f$ for $M_h\!=\!120$~GeV, at both
the Tevatron and the LHC.}
\label{fg:mudep1b}
\end{center}
\end{figure}

\begin{figure}[t]
\begin{center}
\begin{tabular}{rl}
\includegraphics[bb=150 500 430 700,scale=0.70]{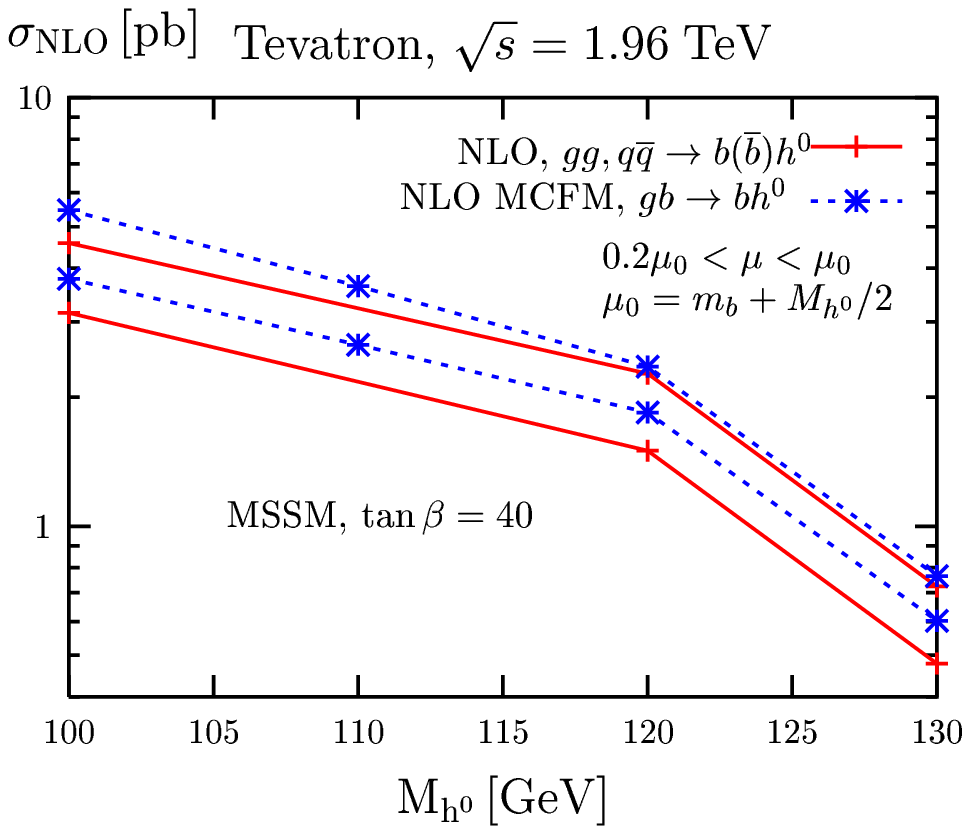} & 
\includegraphics[bb=150 500 430 700,scale=0.70]{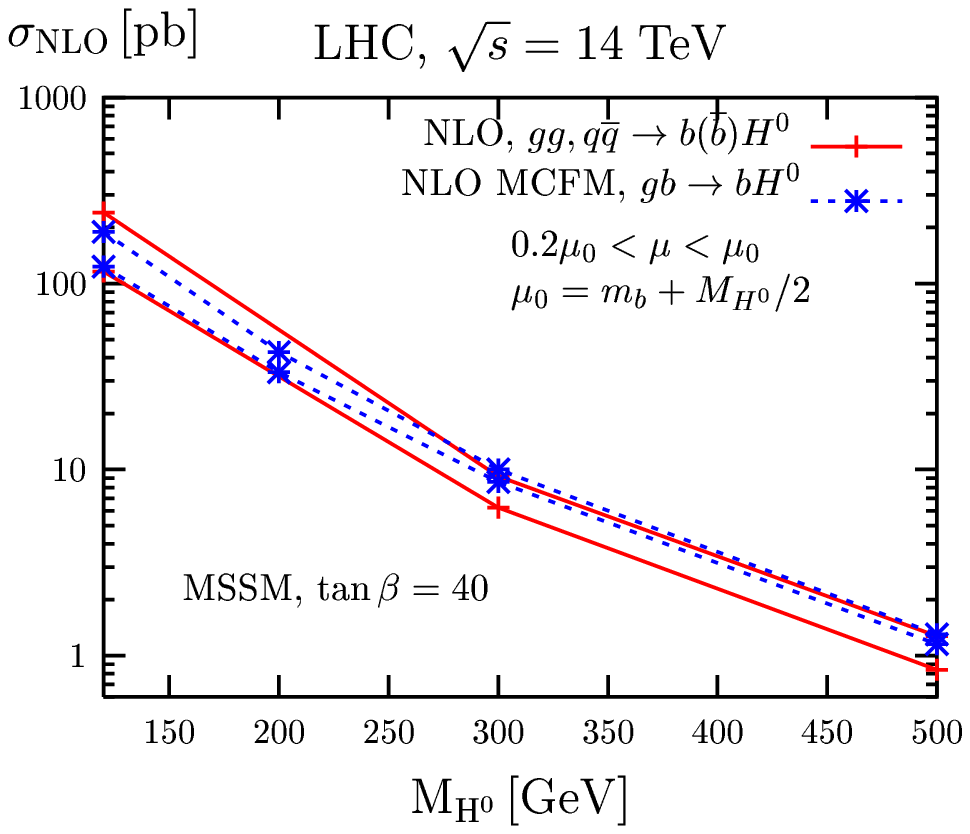}
\end{tabular}
\vspace*{8pt}
\caption[]{Total NLO cross section in the MSSM for $pp,p\bar{p}\rightarrow
  b(\bar{b}) h$ production at the Tevatron and the LHC as a function
  of $M_{h^0,H^0}$. We varied $\mu_r$ and $\mu_f$ independently to
  obtain the uncertainty bands, as explained in the text.  The solid
  curves correspond to the 4FNS, the dashed curves to the 5FNS.}
\label{fg:mhdep1b}
\end{center}
\end{figure}

If a single bottom quark is tagged then the final state is $bh$ or
$\bar{b}h$.  A recent Tevatron study~\cite{Avto:2005} used the search
for neutral MSSM Higgs bosons in events with three bottom quarks in
the final state ($bh+\bar{b}h$ production with $h\rightarrow
b\bar{b}$) to impose limits on the $\tan\beta$ and $M_{A^0}$ parameter
space.  The effect of the NLO corrections in the 4FNS are illustrated
in Fig.~\ref{fg:mudep1b} where we show the SM results for the LO and
NLO total production rates~\cite{Dawson:2004tl}. We see a strong
reduction of the scale dependence at NLO. In Fig.~\ref{fg:mhdep1b} we
compare the rates obtained in the MSSM for $\tan\beta=40$ for the
single $b$-tagged events in the 4FNS and 5FNS schemes, where we varied
separately the renormalization and factorization scales from $0.2
\mu_0$ to $\mu_0$, with $\mu_0\!=\!m_b+M_h/2$. These are obtained
by rescaling the SM results of Ref.~\cite{Dawson:2004tl} according to
Eq.~(\ref{eq:rescale}) to the MSSM
scenario discussed in this review.  Note that the resulting bands only
give an estimate of the theoretical uncertainty due to residual scale
dependence, but do not consider other uncertainties such as
PDF uncertainties. A discussion of the PDF
uncertainties can be found in Section~\ref{sec:pdfs}.  As demonstrated
in Fig.~\ref{fg:mhdep1b} the theoretical predictions in the 4FNS and
5FNS are fully compatible and, thus, either calculation 
can be used in the experimental analyses. The theoretical uncertainty
in both schemes is still dominated by the residual scale dependence
and the PDF uncertainty.

\begin{figure}[t]
\begin{center}
\begin{tabular}{cc}
\includegraphics[bb=140 450 370 725,scale=0.7]{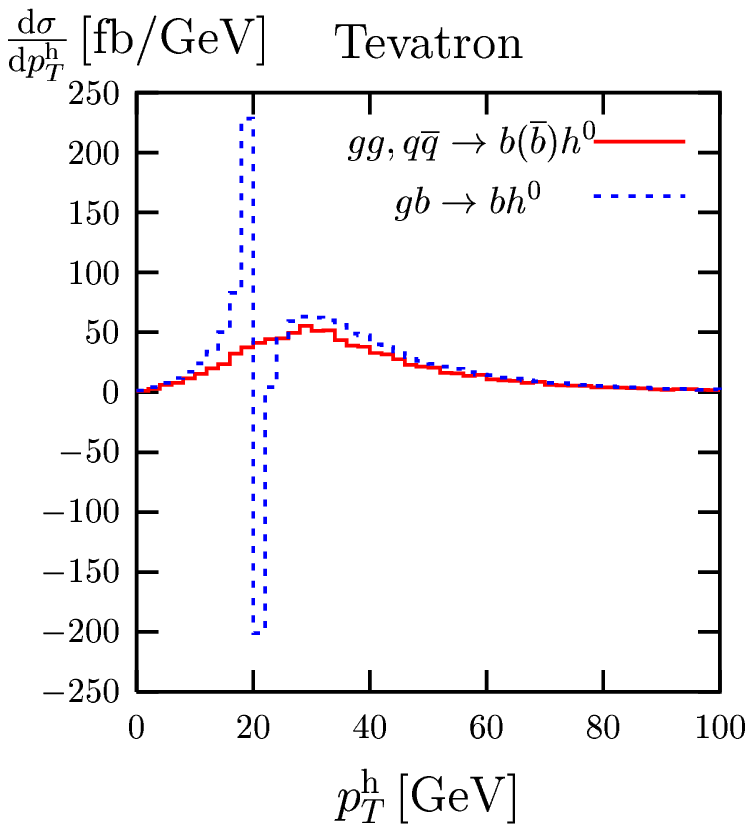}&
\includegraphics[bb=100 455 370 725,scale=0.7]{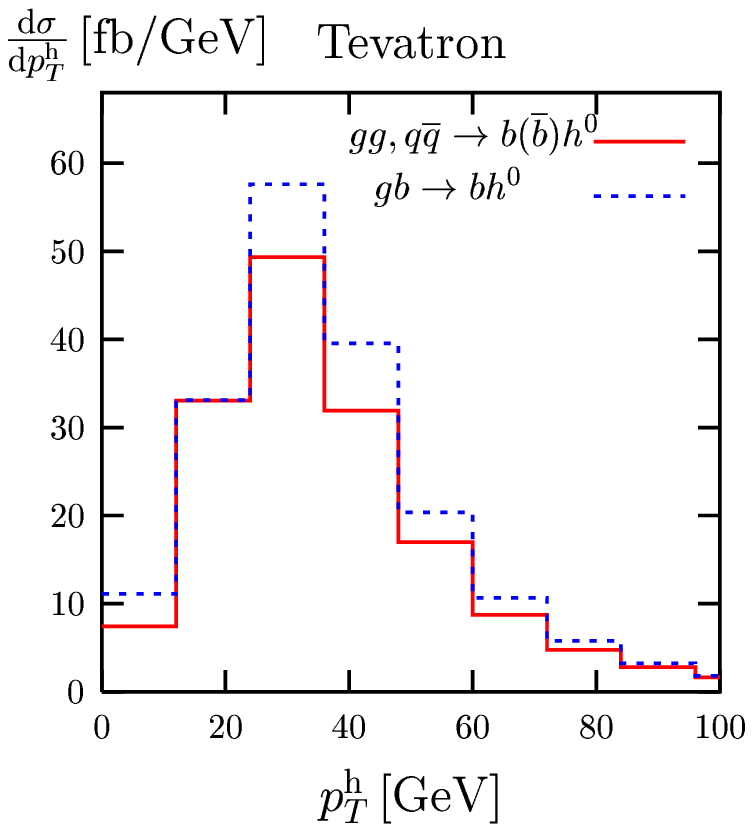}\\
\includegraphics[bb=140 455 370 710,scale=0.7]{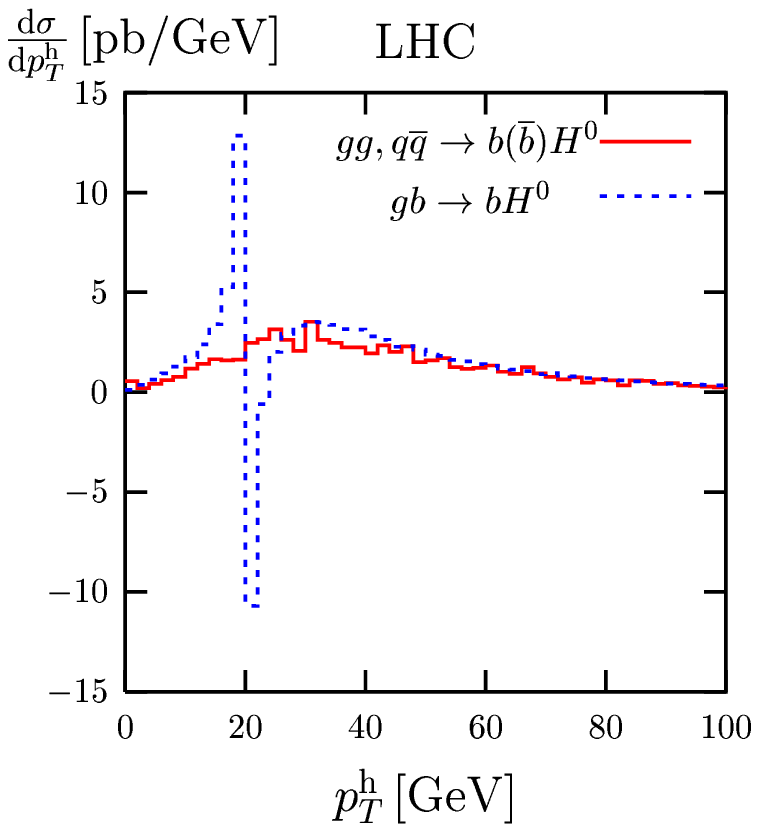}&
\includegraphics[bb=100 455 370 710,scale=0.7]{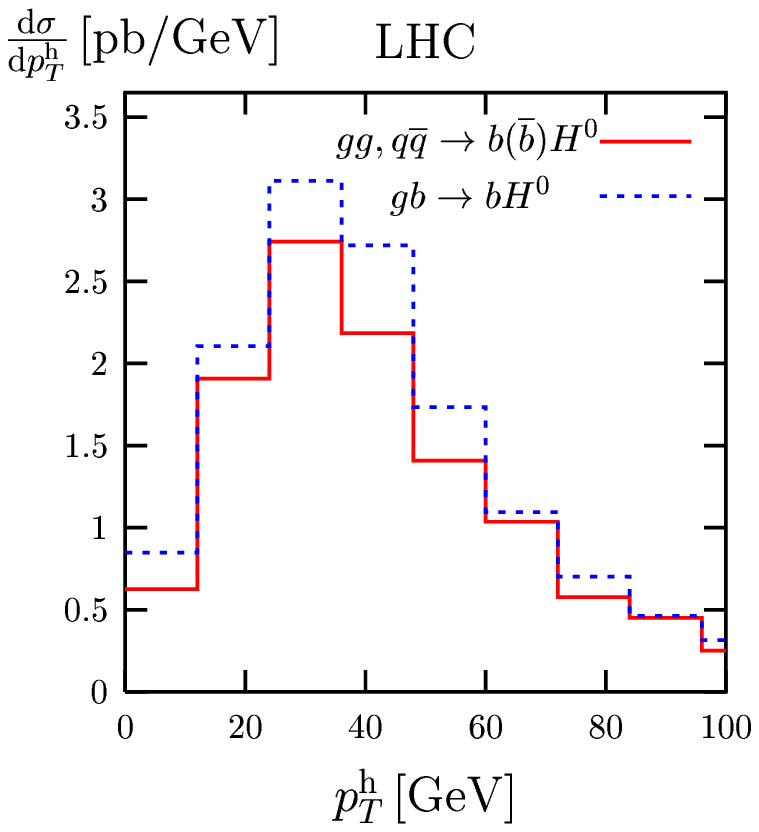}
\end{tabular}
\vspace*{8pt}
\caption[]{$d \sigma/dp_T^h$ in the MSSM at the Tevatron and the LHC for
$M_{h^0,H^0}\!=\!120$~GeV and $\mu_r\!=\!\mu_F\!=\!\mu_0/2$ for single
$b$-tag events.  We show the NLO results in the 4FNS (solid) and 5FNS
(dashed), using two different bin sizes, 2~GeV (left) and 12~GeV
(right).}
\label{fg:dsdpt}
\end{center}
\end{figure}

\begin{figure}[t]
\begin{center}
\begin{tabular}{rl}
\includegraphics[bb=150 500 430 700,scale=0.7]{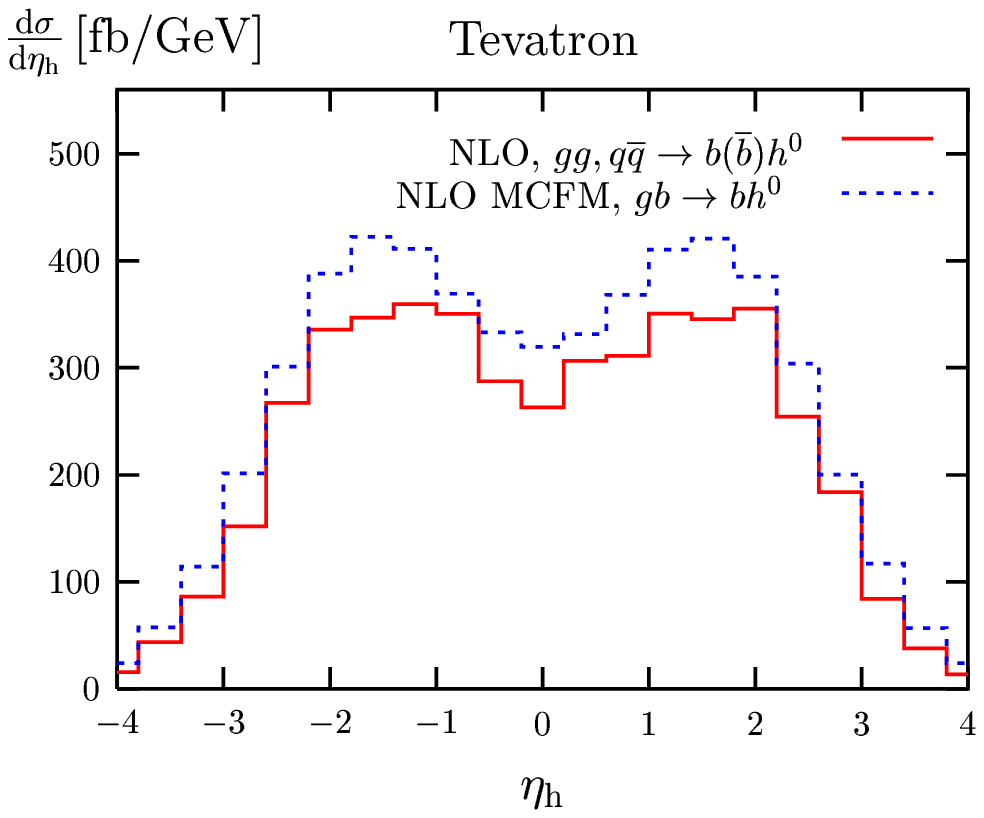} & 
\includegraphics[bb=150 500 430 700,scale=0.7]{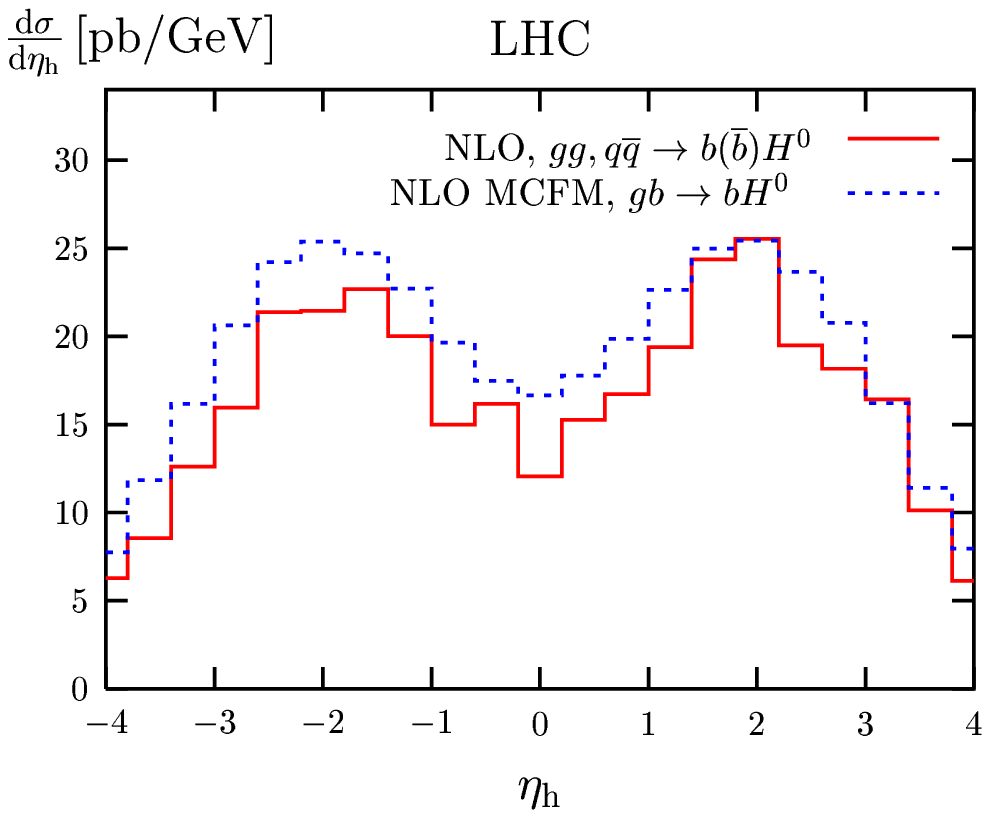}
\end{tabular}
\vspace*{8pt}
\caption[]{$d\sigma/d\eta_h$ in the MSSM at the Tevatron and the LHC for
  $M_{h^0,H^0}\!=\!120$~GeV and $\mu_r\!=\!\mu_f\!=\!\mu_0/2$ for single
  $b$-tag events.  We show the NLO results in the 4FNS (solid) and
  5FNS (dashed).}
\label{fg:dsdeta}
\end{center}
\end{figure}

In comparing the four and five flavor number schemes it is
particularly interesting to compare the kinematic distributions.  In
Figs.~\ref{fg:dsdpt}-\ref{fg:dsdeta}, we compare the results for the
transverse momentum and pseudorapidity distributions of the light and
heavy MSSM Higgs bosons, $h^0$ and $H^0$, in both the 4FNS and 5FNS,
at the Tevatron and the LHC. We see, in general, good agreement
between the two schemes within their respective theoretical
uncertainties, except in regions of kinematic
boundaries. This is particularly dramatic in the $p_T^h$ distributions
in the 5FNS where, around $p_T^h\!\simeq\!20$~GeV, a kinematic
threshold induced in $bg\rightarrow bh$ by the cut on the $p_T$ of the
bottom quark causes
the 5FNS NLO calculation to be unreliable. This instability can be
reabsorbed by using a larger bin size (see inlays), and could
therefore be interpreted as a sort of theoretical \emph{resolution}
for the 5FNS.  The instabilities could be removed by a systematic
resummation of threshold
corrections~\cite{Catani:1997xc,DelDuca:2003uz}, which is not
implemented in {\sc MCFM}.

\subsection{Results for exclusive $b\bar b h$ production}
\label{subsec:2btag}
\begin{figure}[t]
\begin{center}
\begin{tabular}{rl}
\includegraphics[bb=40 30 520 430,scale=0.40]{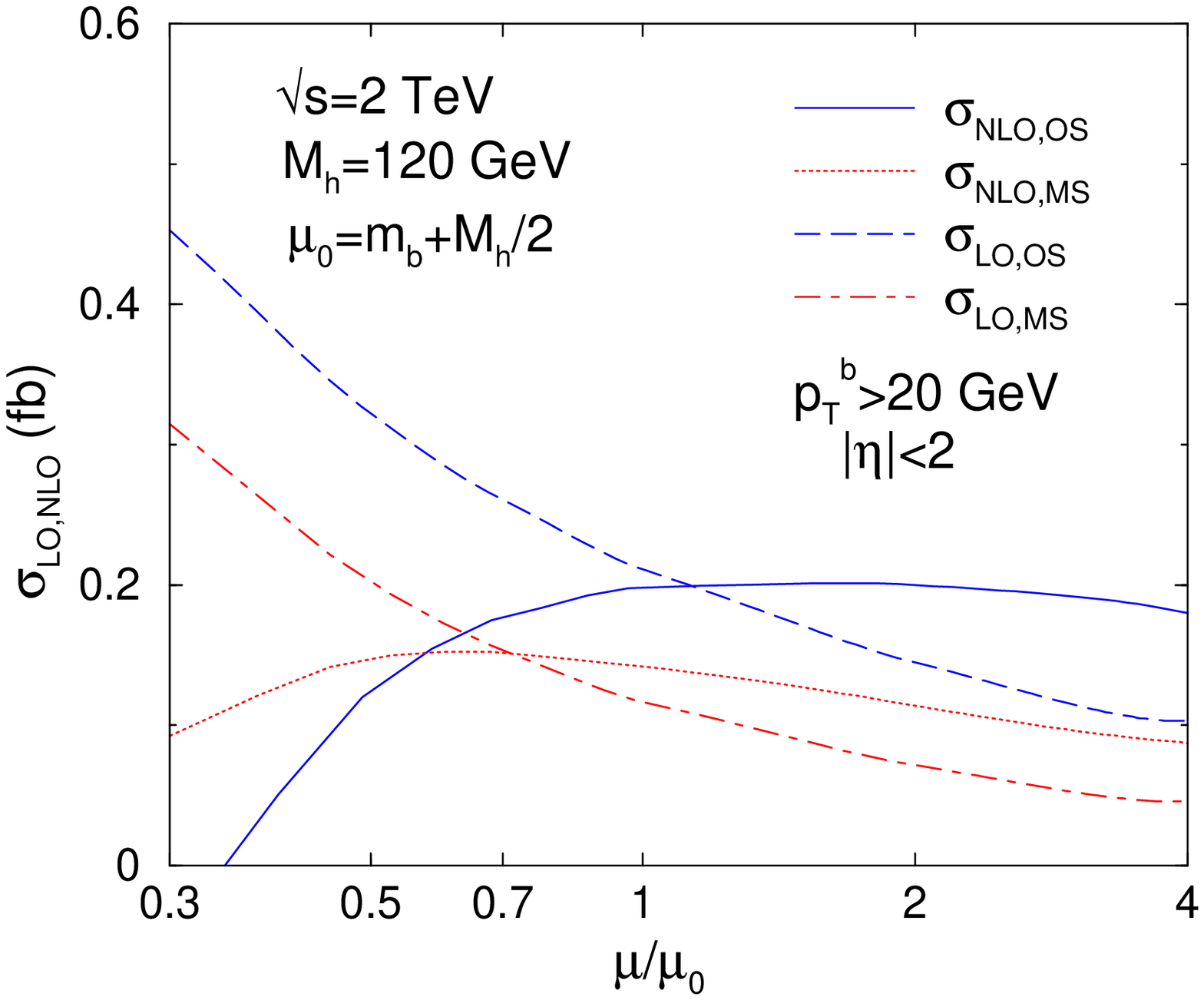} &
\includegraphics[bb=40 30 520 430,scale=0.40]{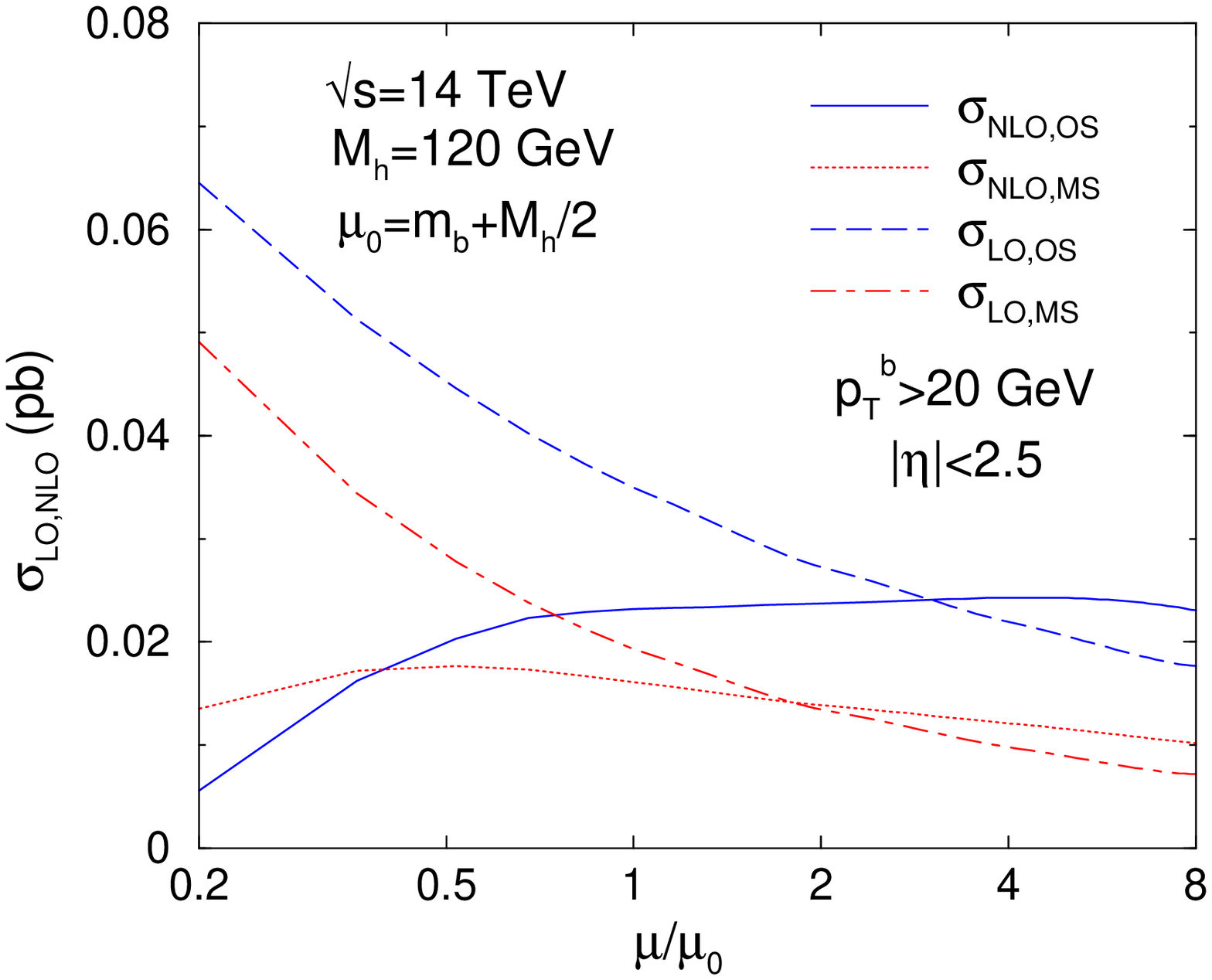}
\end{tabular}
\vspace*{-0.4cm}
\caption[]{Total LO and NLO cross sections in the SM for the two 
  $b$-tag process, $pp,p\bar{p}\rightarrow b\bar{b}h$, in the 4FNS as
  a function of $\mu\!=\!\mu_r\!=\!\mu_f$ for $M_h\!=\!120$~GeV, at
  both the Tevatron and the LHC. Results are shown for both the
  $\overline{MS}$ and $OS$ renormalization scheme for the bottom Yukawa coupling (see discussion in
  Section~\ref{sec:renorm}).}
\label{fg:mudep}
\end{center}
\end{figure}

Finally, we discuss the fully exclusive LO and NLO cross sections for
$b\bar{b}h$ production, where both the outgoing $b$ and $\bar{b}$
quarks are identified.  The results of Ref.~\cite{Dawson:2003ex} that
are presented here have been obtained by using the CTEQ5 set of PDFs.
We have checked that the conclusions of the numerical discussion
presented here do not change when using the CTEQ6 set of PDFs.  In
Fig.~\ref{fg:mudep} we show, for $M_h\!=\!120$~GeV, the dependence of
the LO and NLO total cross sections, calculated in the SM and in the
4FNS, on the arbitrary renormalization/factorization scale $\mu$ (with
$\mu_r\!=\!\mu_f\!=\!\mu$). The curves labelled "$OS$" ("$MS$") use
the on-shell (${\overline{MS}}$) scheme for the bottom Yukawa
coupling.  The NLO results have significantly less sensitivity to the
scale choice, both in the $OS$ and in the $\overline{MS}$
schemes. However, while in the $OS$ scheme the plateau region, or
region of least sensitivity to the renormalization/factorization
scale, is around $\mu\!=\!\mu_0$, in the $\overline{MS}$ scheme it is
shifted towards $\mu\!=\!\mu_0/2$, ($\mu_0\!=\!m_b+M_h/2$). It is
interesting to note, that the scale choice $\mu=M_h/4$ is supported by
theoretical studies~\cite{Harlander:2003ai} and corresponds to the
point where the NNLO rate for inclusive Higgs production from bottom
quarks is the same as the NLO rate~\cite{Boos:2003yi,Maltoni:2003pn}.
Fig.~\ref{fg:mudep} shows how the $\overline{MS}$ scheme gives a
perturbative cross section that is better behaved over a broader
range of scales, and therefore preferable.  Conservatively, one could
interpret the difference between the two plateau regions as an
additional 10-20\% theoretical error.

\begin{figure}[t]
\begin{center}
\begin{tabular}{rl}
\includegraphics[bb=150 500 430 700,scale=0.6]{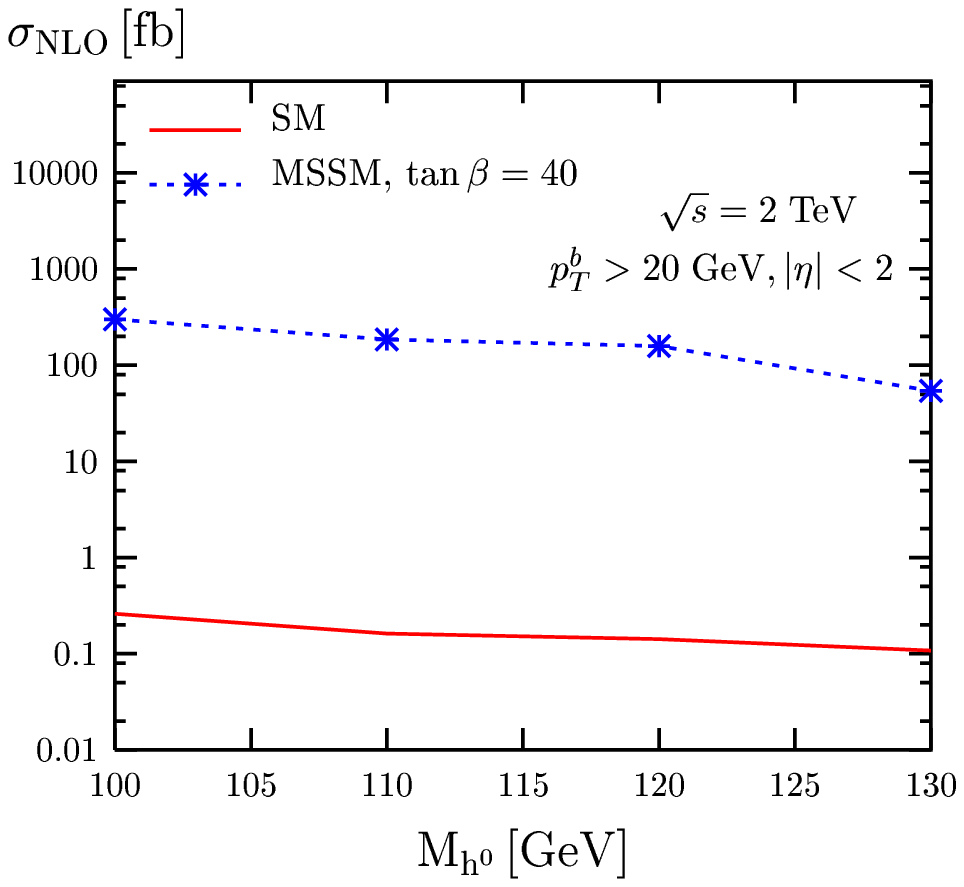} & 
\includegraphics[bb=150 500 430 700,scale=0.6]{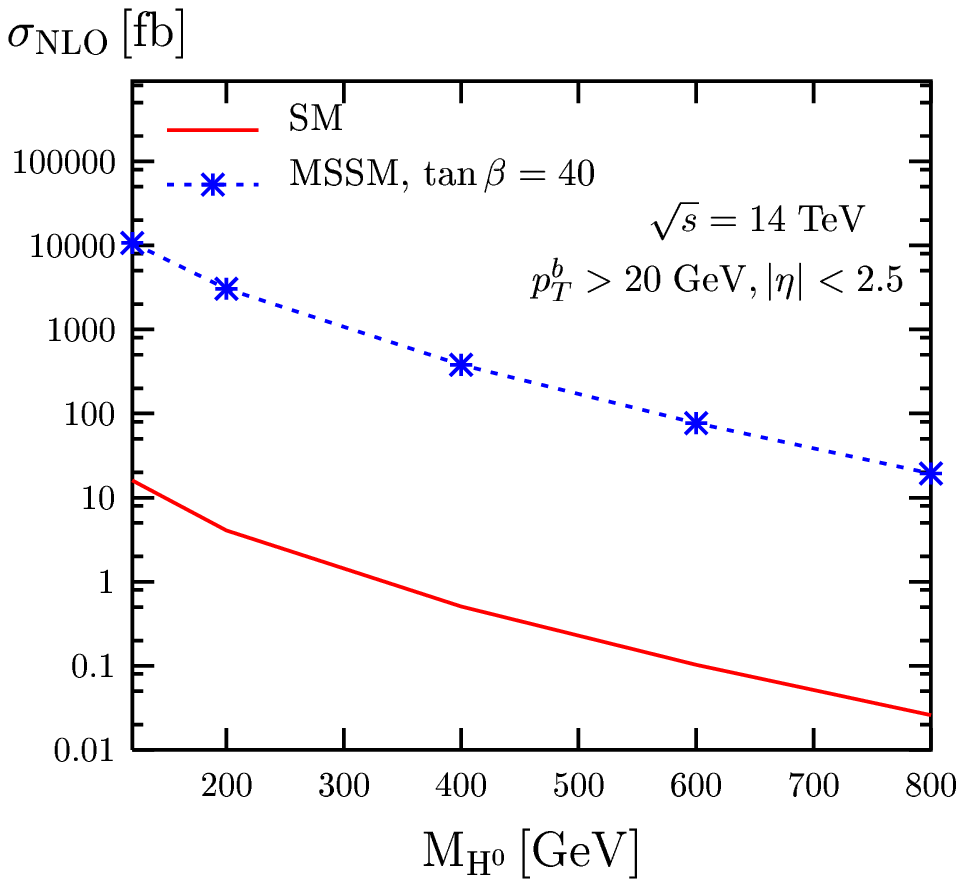}
\end{tabular}
\vspace*{8pt}
\caption[]{Total NLO cross sections in the SM and MSSM for the
  two $b$-tag process, $pp,p\bar{p}\rightarrow b\bar{b}h$, in the 4FNS
  for the central scale $\mu_r\!=\!\mu_f\!=\!\mu_0$. Shown is the
  dependence on the light and heavy MSSM Higgs boson masses at the
  Tevatron and the LHC, respectively.}
\label{fg:mhdep}
\end{center}
\end{figure}

In Fig.~\ref{fg:mhdep} we compare the production of the SM Higgs boson
with that of the neutral scalar Higgs bosons of the MSSM and again
observe a significant enhancement of the rate in the MSSM for large
$\tan\beta$. The MSSM results have been obtained from the SM results
as described in Eq.~\ref{eq:rescale} by using the rescaling factors of
Tables~\ref{tab:ratiolight}, \ref{tab:ratioheavy} for $\tan\beta=40$.

\begin{figure}[t]
\begin{center}
\includegraphics[bb=150 500 430 700,scale=0.6]{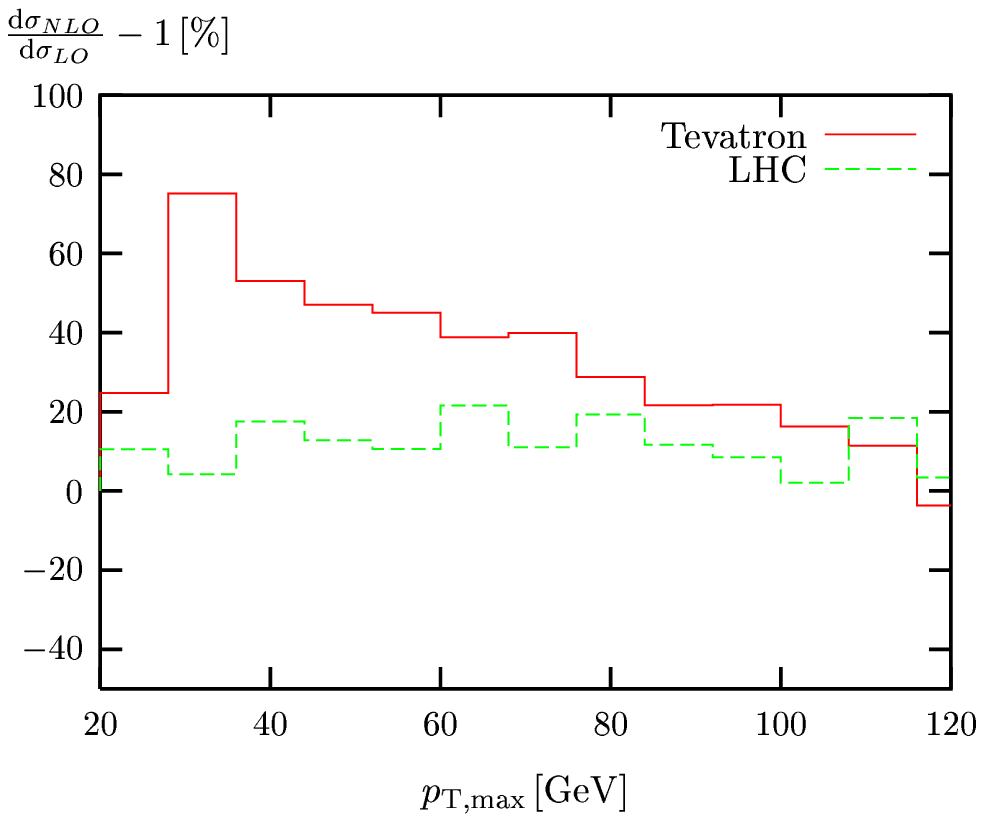} 
\includegraphics[bb=150 500 430 700,scale=0.6]{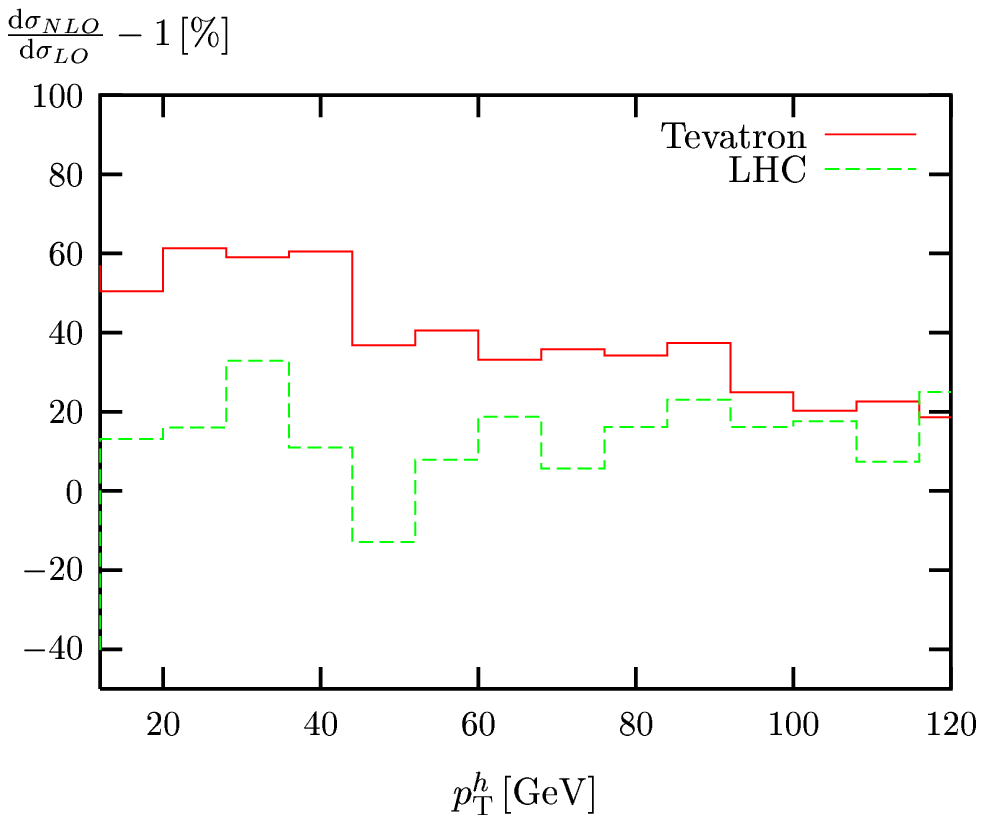} 
\caption[]{The relative corrections $d\sigma_{\sss NLO}/d \sigma_{\sss
LO}-1$ for the $p_T$ distribution of the bottom or anti-bottom quark
with the largest $p_T$ ($p_{T,max}$) (left) and of the SM Higgs boson
($p_{T}^h$) (right) to $b\bar b h$ production in the SM at the
Tevatron (with $\sqrt{s}\!=\!2$~TeV and $\mu\!=\!2 \mu_0$) and the LHC
(with $\sqrt{s}\!=\!14$~TeV and $\mu\!=\!4 \mu_0$).}
\label{fg:bbh_ptrel}
\end{center}
\end{figure}

\begin{figure}[t]
\begin{center}
\includegraphics[bb=150 500 430 700,scale=0.6]{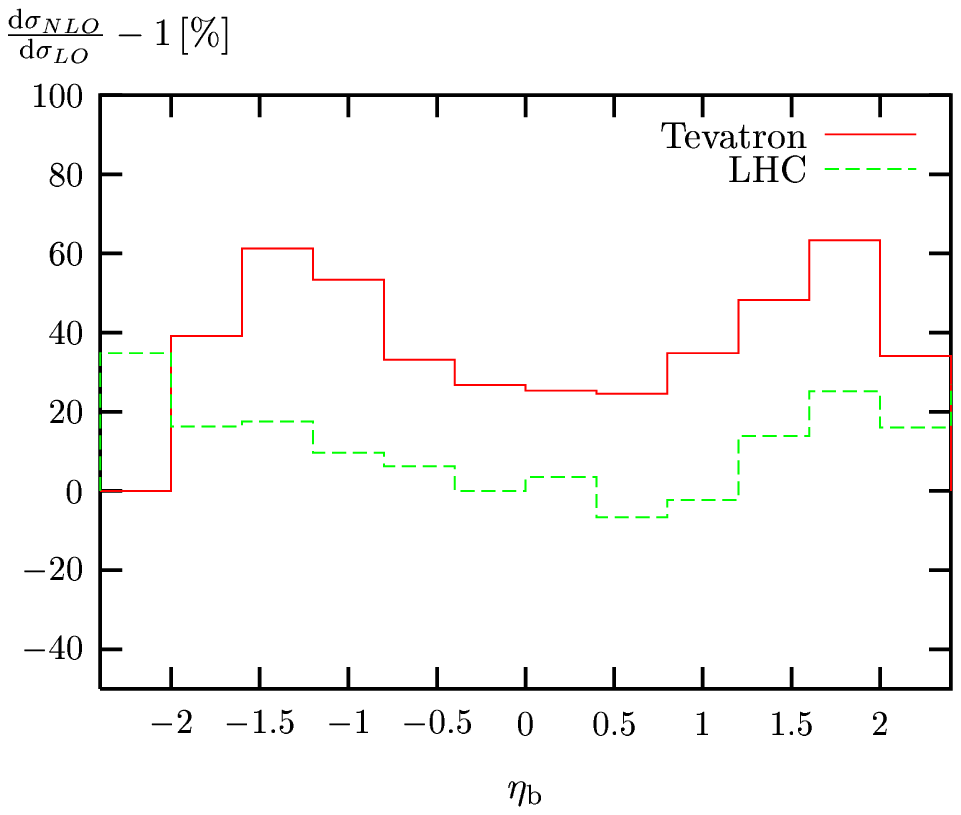} 
\includegraphics[bb=150 500 430 700,scale=0.6]{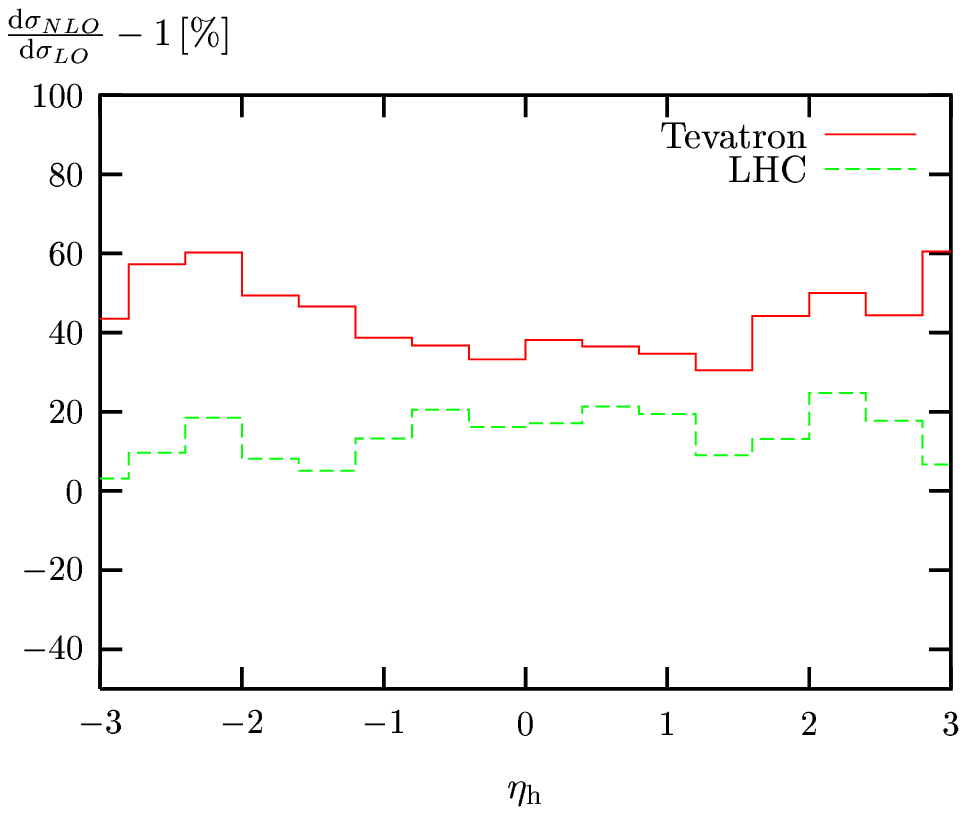} 
\caption[]{The relative corrections $d\sigma_{\sss NLO}/d \sigma_{\sss
LO}-1$ for the $\eta$ distribution of the bottom quark $\eta_b$ (left)
and of the SM Higgs boson ($\eta_{h}$) (right) to $b\bar b h$
production in the SM at the Tevatron (with $\sqrt{s}\!=\!2$~TeV and
$\mu\!=\!2 \mu_0$) and the LHC (with $\sqrt{s}\!=\!14$~TeV and
$\mu\!=\!4 \mu_0$).}
\label{fg:bbh_etarel}
\end{center}
\end{figure}

Finally, in Figs.~\ref{fg:bbh_ptrel}, \ref{fg:bbh_etarel} we
illustrate the impact of NLO QCD corrections on the transverse
momentum and pseudorapidity distribution of the SM Higgs boson and the
bottom quark by showing the relative correction,
$d\sigma_{NLO}/d\sigma_{LO}-1$ (in percent).  
For the renormalization/factorization scale we choose $\mu=2 \mu_0$
at the Tevatron and $\mu=4 \mu_0$ at the LHC. These two scale choices are
well within the plateau regions where the NLO cross sections vary the
least with the value of $\mu$.
As can be seen, the NLO
QCD corrections can considerably affect the shape of kinematic
distributions, and their effect cannot be obtained from simply
rescaling the LO distributions with a K-factor of $\sigma_{\sss
NLO}/\sigma_{\sss LO}\!=\!1.38\pm 0.02$ (Tevatron, $\mu\!=\!2\mu_0$)
and $\sigma_{\sss NLO}/\sigma_{\sss LO}\!=\!1.11\pm 0.03$ (LHC,
$\mu\!=\!4\mu_0$). The errors only include the statistical
uncertainty from the Monte Carlo integration. 
The kinematic distributions have been calculated
within the SM and using the $OS$ scheme, but we see a similar
behavior when using the $\overline{MS}$ bottom quark Yukawa coupling
or consider the MSSM case.

\section{PDF Uncertainties}
\label{sec:pdfs}
Besides the residual renormalization/factorization scale dependence
after the NLO corrections have been included, another major source of
theoretical uncertainty for cross section predictions at hadron
colliders comes from the Parton Distribution Functions.
Unfortunately, PDFs are plagued by uncertainties either from the
non-perturbative starting distributions used to fit the data or from
the perturbative DGLAP evolution to the higher energies relevant at
hadron colliders~\cite{Gribov:1972,Altarelli:1977zs,Dokshitzer:1977}.

Recently, several collaborations have introduced new schemes which
allow an estimate of theoretical uncertainties on physical observables
due to the uncertainty in the PDFs.  Here, we focus on the scheme
introduced by the CTEQ collaboration based on the Hessian matrix
method~\cite{CTEQ:2002} and study the uncertainties of 
semi-inclusive total $b\bar bh$ production rates that are induced by
the uncertainties in the PDFs.  The details of this method are beyond
the scope of this review, however, we give a brief explanation below.
First, the nominal set of PDFs (e.g. CTEQ6) is constructed by fitting
a {\it non-perturbative core equation} to data from low-energy
experiments designed to measure PDFs.  The core equation, in the
method used by CTEQ, is parameterized by 20 independent parameters
which are dialed to fit the data.  Once the nominal set is fixed, the
20 parameters are then varied in a well-defined manner to produce an
additional 40 sets of PDFs.  These sets serve as a \emph{map} of the
neighborhood around the nominal fit to the data.  Indeed, one can then
use the 40 sets to estimate the uncertainty from the PDFs on a
physical observable in the following way~\cite{Ferrag:2003}
\footnote{We have also performed this analysis using the PDF sets of
the MRST collaboration~\cite{Martin:2002aw}.  These sets are made up
of 30 sets in addition to the nominal fit and, hence, map less of the
neighborhood around the global minimum.  This results in smaller
spread uncertainties than the CTEQ analysis.  Therefore, we only show
results using the CTEQ sets and quote these results as an upper limit
of the uncertainty from PDFs.}:
first, the central value cross section $\sigma_0$ is calculated using
the global minimum PDF (i.e. CTEQ6M).  The calculation of the cross
section is then performed with the additional 40 sets of PDFs to
produce 40 different predictions, $\sigma_i$.  For each of these, the
deviation from the central value is calculated to be
$\Delta\sigma_i^{\pm} = |\sigma_i-\sigma_0|$ when ${\sigma_i}_{<}^{>}
\sigma_0$.  Finally, to obtain the uncertainties due to the PDFs the
deviations are summed quadratically as $\Delta\sigma^{\pm} = \sqrt{
\sum_i {\Delta\sigma^{\pm}_{i}}^{2}}$ and the cross section including
the theoretical uncertainties arising from the PDFs is quoted as
$\sigma_0|^{+\Delta\sigma^{+}}_{-\Delta\sigma^{-}}$.  Recently, similar 
analyses have been performed for the dominant SM Higgs production 
modes~\cite{Djouadi:2003pd} and for the fully inclusive Higgs production
modes in both the SM ($gg \to h$) and the MSSM 
($b\bar{b} \to h$)~\cite{Belyaev:2005nu}.
\begin{figure}[t]
\begin{center}
\includegraphics[scale=0.3,angle=-90]{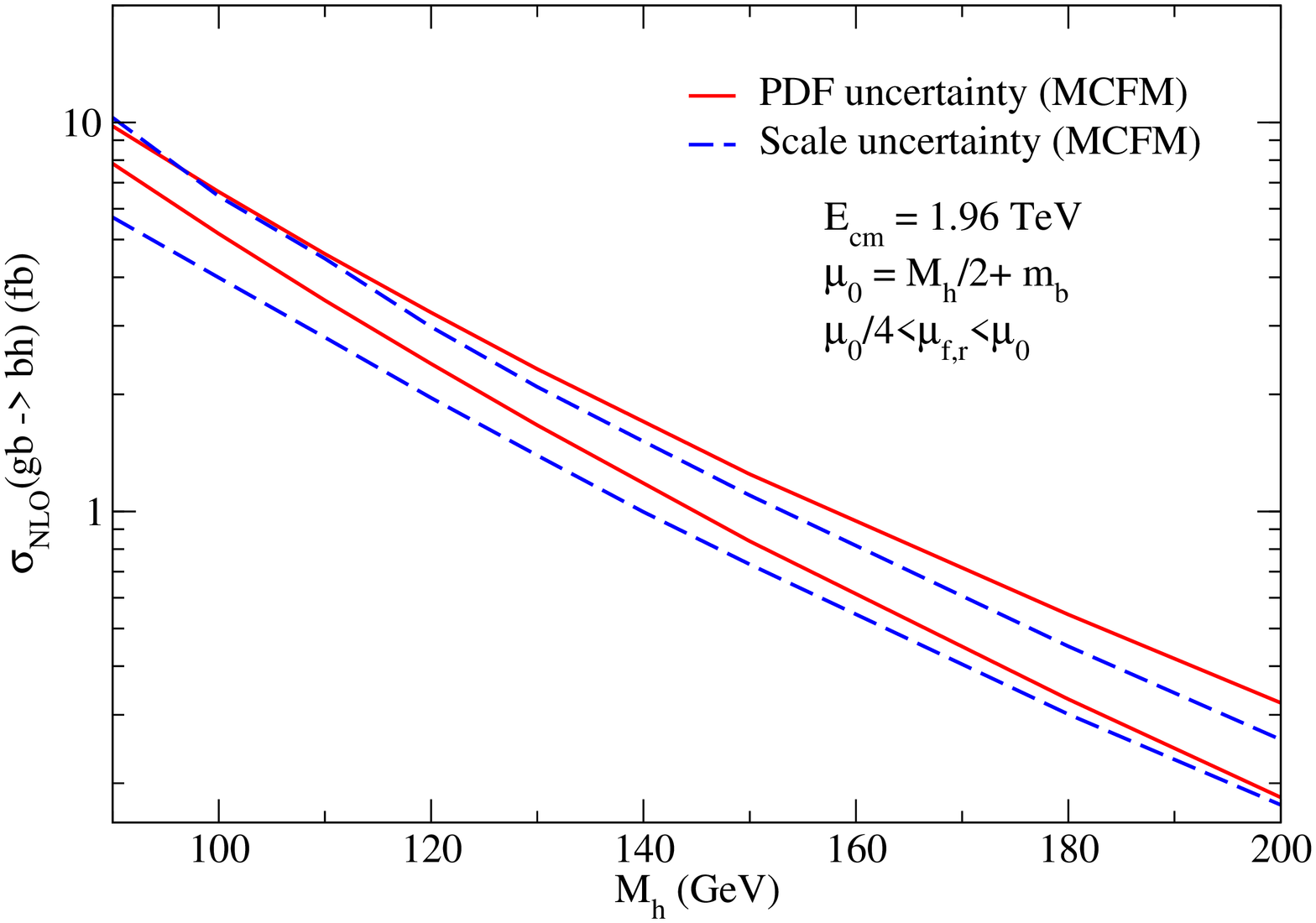}
\includegraphics[scale=0.3,angle=-90]{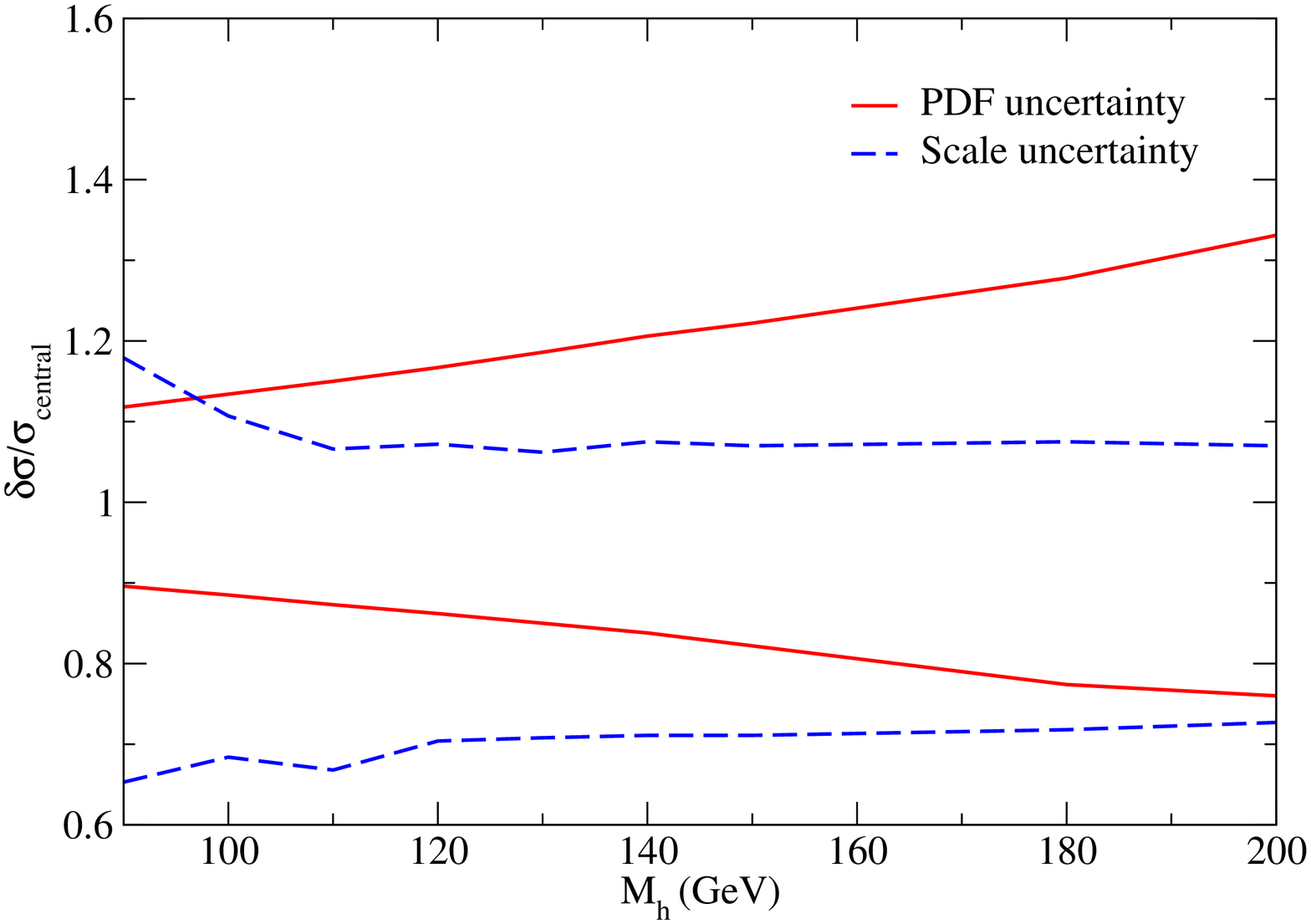}
\caption[]{Comparison between theoretical uncertainties due to scale
dependence and uncertainties arising from the PDFs at the Tevatron. 
In the bottom plot, both uncertainty bands have been normalized to the 
central value of the total cross section $\sigma_0$.}
\label{fg:tevPDF}
\end{center}
\end{figure}
\begin{figure}[t]
\begin{center}
\includegraphics[scale=0.3,angle=-90]{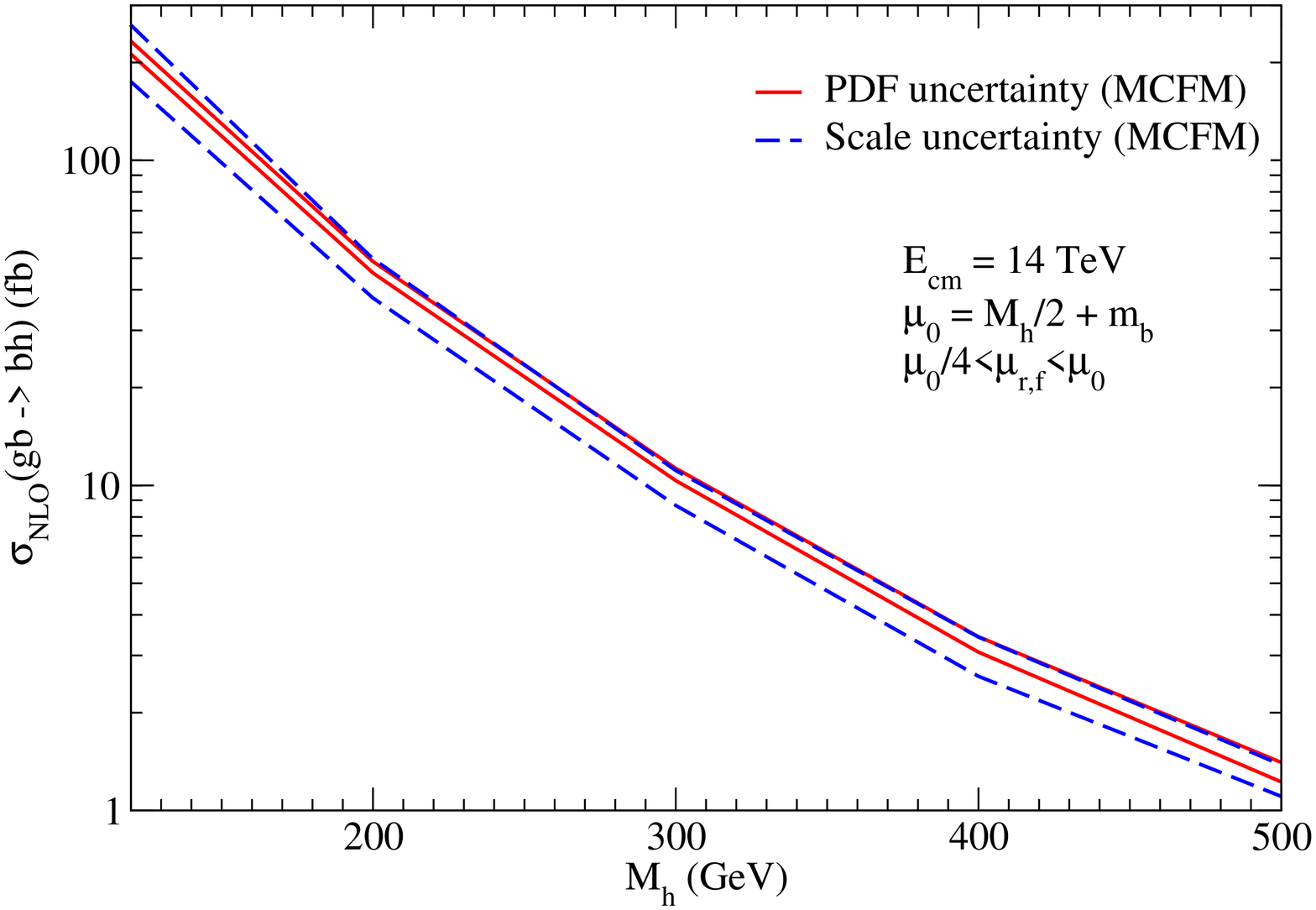}
\includegraphics[scale=0.3,angle=-90]{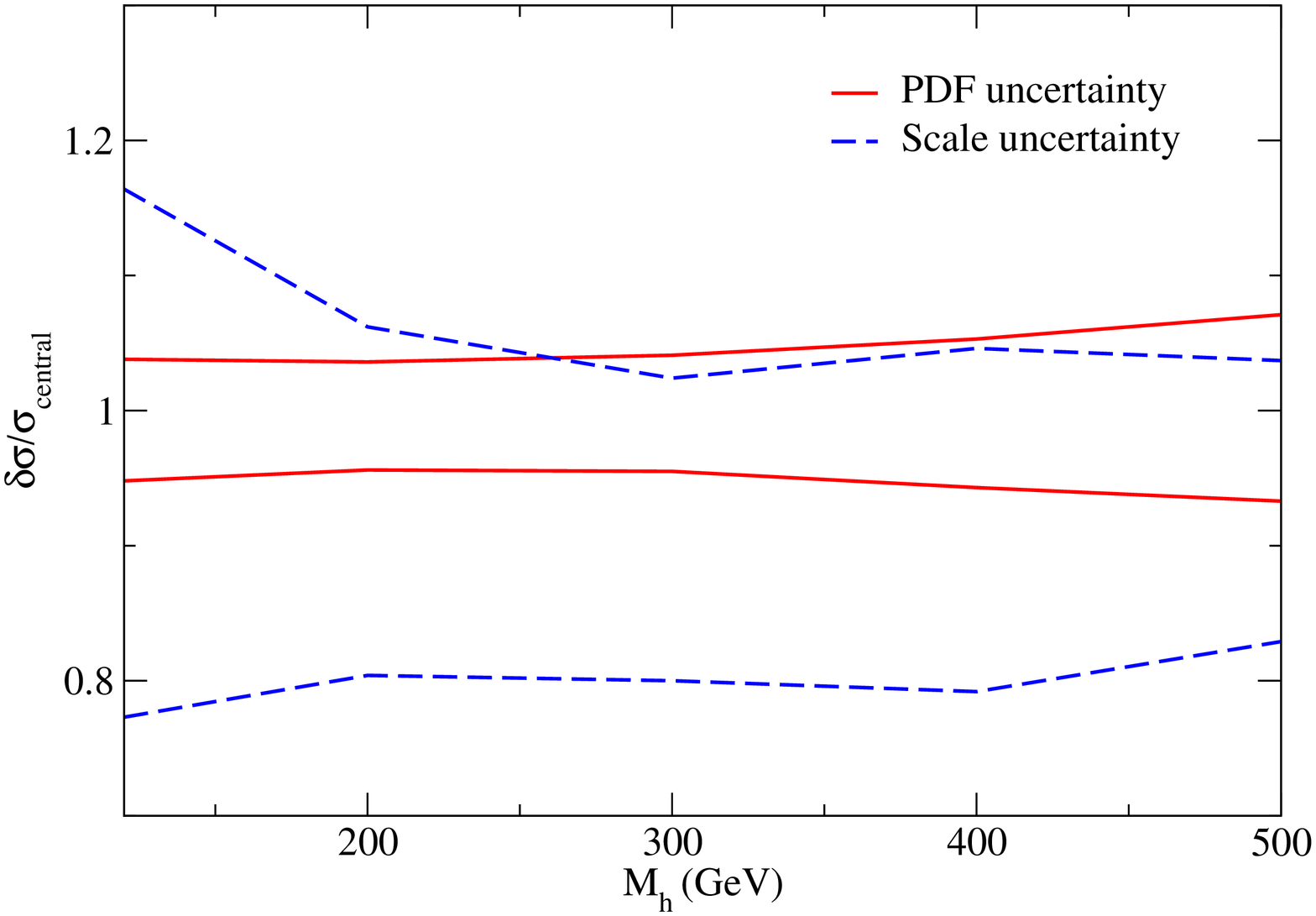} 
\caption[]{Comparison between theoretical uncertainties due to scale
dependence and uncertainties arising from the PDFs at the LHC.
In the bottom plot, both uncertainty bands have been normalized to the 
central value of the total cross section $\sigma_0$.}
\label{fg:lhcPDF}
\end{center}
\end{figure}
In Figs.~\ref{fg:tevPDF} and~\ref{fg:lhcPDF} we plot the total SM NLO
cross section for $bg\rightarrow bh$ obtained with
MCFM~\cite{MCFM:2004} at the Tevatron and LHC respectively.  Here, we
compare the spread uncertainties from residual scale dependence and
the PDFs both for the total cross section (top) and the total cross
section normalized to the central value calculated with CTEQ6M
(bottom).

From Fig.~\ref{fg:lhcPDF} one can see that, at the LHC, the
theoretical uncertainty is dominated by the residual scale dependence.
Due to the large center of mass (c.o.m.) energy of the LHC, the gluons
and bottom quarks in the initial state have small momentum fraction
($x$) values and, hence, small PDF uncertainties typically in the
5-10\% range.

In contrast, due to the smaller c.o.m. energy, the PDF uncertainties
at the Tevatron (Fig.~\ref{fg:tevPDF}) are comparable and even
larger than the uncertainties due to residual scale dependence over
the full Higgs mass range.  The smaller c.o.m. energy results in
higher-$x$ gluons and bottom quarks in the initial state which
corresponds to large PDF uncertainties in the 10-30\% range.

\begin{figure}[t]
\begin{center}
\includegraphics[scale=0.3,angle=-90]{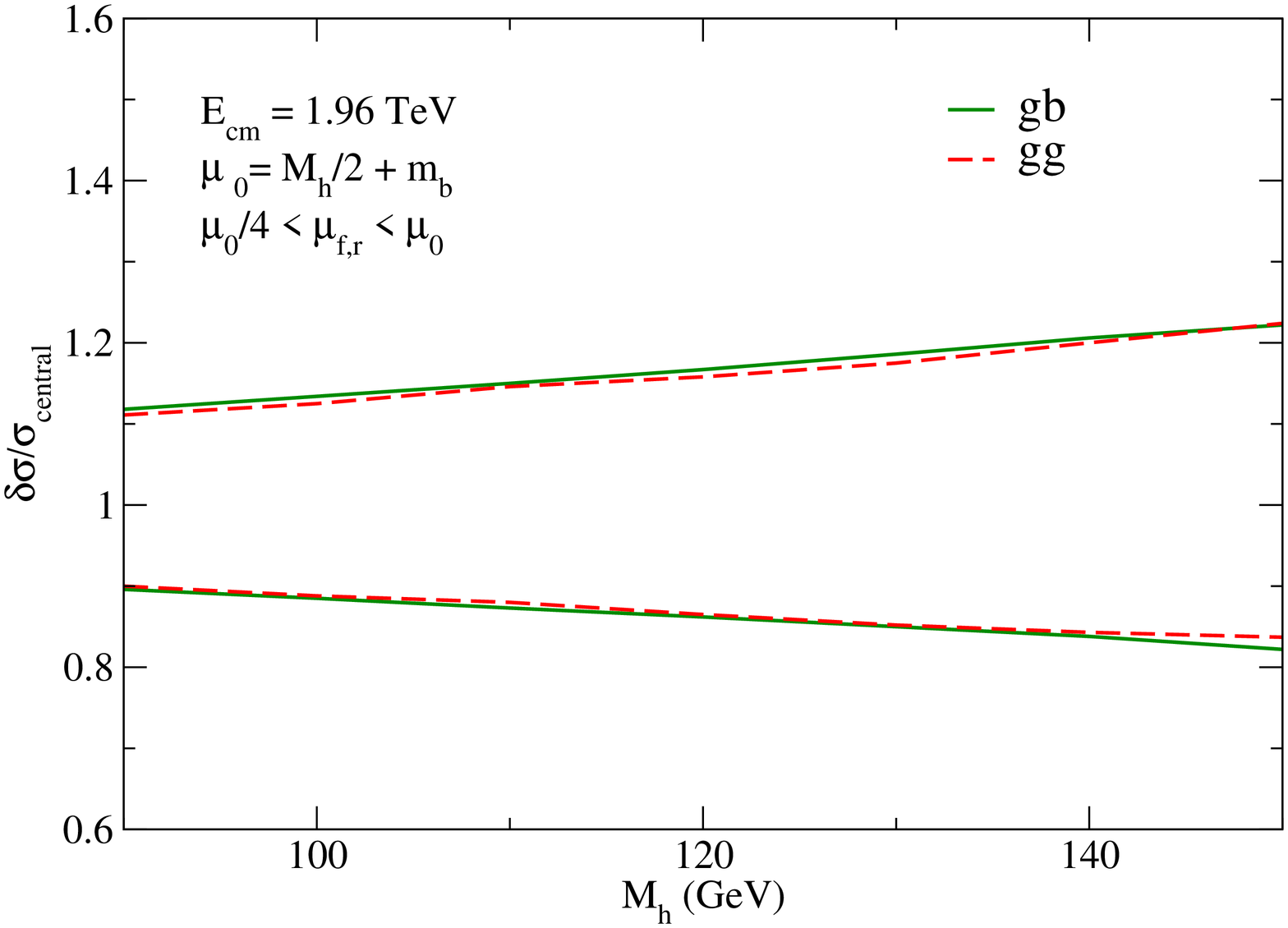}
\includegraphics[scale=0.3,angle=-90]{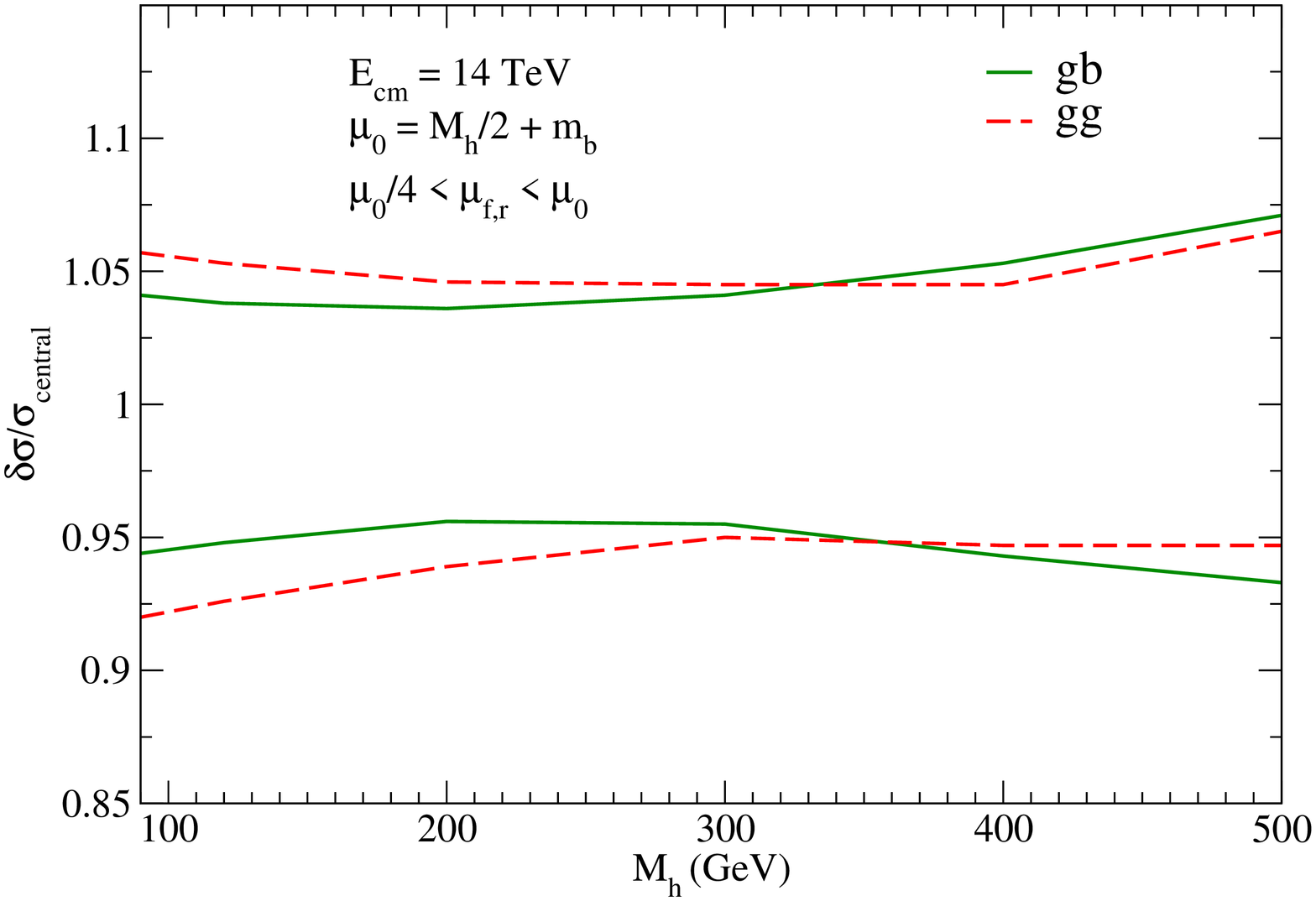} 
\caption[]{Normalized cross sections for Higgs production with one $b$
jet at the Tevatron (top) and the LHC (bottom) showing the uncertainty
from PDFs for both the $gg$ and $bg$ initial states.}
\label{fg:ggvbg}
\end{center}
\end{figure}

Finally, in Fig.~\ref{fg:ggvbg}, we plot the normalized total SM NLO
cross sections of $bg\rightarrow bh$ and $gg\rightarrow b(\bar{b})h$
and compare their respective uncertainties due to the PDFs.  We see
that, at both the Tevatron and the LHC, the PDF uncertainties are
almost identical for both the $gg$ and $bg$ initial states.

\section{Conclusions}
\label{sec:summary}
We discussed the status of theoretical predictions for Higgs boson
production in association with bottom quarks for the cases where zero,
one or both outgoing bottom quarks are identified. These processes are
important Higgs boson discovery modes in models with enhanced bottom
quark Yukawa couplings, such as the MSSM at large values of
$\tan\beta$.  We presented results for the MSSM $b\bar{b}h$
($h=h^0,H^0$) total and differential production rates at
$\tan\beta=40$ and discussed the theoretical uncertainties of the
state-of-the-art predictions due to the residual
renormalization/factorization scale dependence at NLO (NNLO) in QCD
and the uncertainty in the PDFs.  We have shown that the total cross
section for the inclusive case, $pp,p\bar{p}\rightarrow (b\bar{b})h$,
and both the total and differential cross
sections for the semi-inclusive case, $pp,p\bar{p}\rightarrow
b(\bar{b})h$, within a 4FNS and 5FNS are fully compatible within the
existing theoretical errors.  All important Higgs production processes
at hadron colliders are now known at NLO, and some even at NNLO, and
the theoretical uncertainties of the QCD predictions are well under
control.
\section*{Acknowledgments}
We thank S.~Dittmaier, M.~Kr\"amer, and M.~Spira for comparing
results; R.~Harlander and W.~Kilgore for providing us with the results
of their calculation, and J.~Campbell, F.~Maltoni, and
S.~Willenbrock for discussions.  The work of S.D. and C.B.J. (L.R.) 
is supported in part by the U.S. Department of Energy under
grant DE-AC02-98CH10886 (DE-FG02-97ER41022). The work of D.W. 
is supported in part by the National
Science Foundation under grant NSF-PHY-0244875. L.R. and D.W.
would also like to thank the Aspen Center for Physics, where
part of this work was done,
for their hospitality and financial support.


\end{document}